\documentclass{aa}  

\usepackage{graphicx}
\usepackage{txfonts}
\usepackage{bm}
\usepackage{booktabs,caption}

\usepackage{xcolor}

\usepackage{hyperref}

\hypersetup{
  colorlinks   = true,    
  urlcolor     = blue,    
  linkcolor    = blue,    
  citecolor    = blue     
}
\usepackage{etoolbox}

\makeatletter

\pretocmd{\NAT@citex}{%
  \let\NAT@hyper@\NAT@hyper@citex
  \def\NAT@postnote{#2}%
  \setcounter{NAT@total@cites}{0}%
  \setcounter{NAT@count@cites}{0}%
  \forcsvlist{\stepcounter{NAT@total@cites}\@gobble}{#3}}{}{}
\newcounter{NAT@total@cites}
\newcounter{NAT@count@cites}
\def\NAT@postnote{}

\def\NAT@hyper@citex#1{%
  \stepcounter{NAT@count@cites}%
  \hyper@natlinkstart{\@citeb\@extra@b@citeb}#1%
  \ifnumequal{\value{NAT@count@cites}}{\value{NAT@total@cites}}
    {\ifNAT@swa\else\if*\NAT@postnote*\else%
     \NAT@cmt\NAT@postnote\global\def\NAT@postnote{}\fi\fi}{}%
  \ifNAT@swa\else\if\relax\NAT@date\relax
  \else\NAT@@close\global\let\NAT@nm\@empty\fi\fi
  \hyper@natlinkend}
\renewcommand\hyper@natlinkbreak[2]{#1}

\patchcmd{\NAT@citex}
  {\ifNAT@swa\else\if*#2*\else\NAT@cmt#2\fi
   \if\relax\NAT@date\relax\else\NAT@@close\fi\fi}{}{}{}
\patchcmd{\NAT@citex}
  {\if\relax\NAT@date\relax\NAT@def@citea\else\NAT@def@citea@close\fi}
  {\if\relax\NAT@date\relax\NAT@def@citea\else\NAT@def@citea@space\fi}{}{}

\makeatother

\begin{document} 

   \title{Growth of Ex-situ Diffuse Intragroup Light in Simulated Galaxy Groups}

  \author{B.\ Bilata-Woldeyes
          \inst{1}\fnmsep\thanks{Email: betelehem@iaa.es}, 
          J.\ D.\ Perea\inst{1}
          \and
          J.\ M.\ Solanes\inst{2}
          }

   \institute{Instituto de Astrof\'\i sica de Andaluc\'\i a (IAA–CSIC), Glorieta de la Astronom\'\i a, s/n, E-18008 Granada, Spain
         \and
             Departament de F\'\i sica Qu\`antica i Astrof\'\i sica and Institut de Ci\`encies del Cosmos (ICCUB), Universitat de Barcelona. C.\ Mart\'{\i}  i Franqu\`es, 1, E-08028 Barcelona, Spain
             }
             
   \date{v.1.3 19/Feb/2025. Accepted for publication}

  \abstract
   {Deep surface photometry reveals the presence in a significant fraction of galaxy groups of a faint and diffuse baryonic component permeating the intragroup space. This intragroup light (IGL) is primarily formed by stars that are removed from their host galaxies through gravitational interactions and now drift freely, unbound to any particular galaxy.}
   {We conduct a detailed analysis to investigate how various physical parameters of galaxy groups influence the formation of ex-situ IGL during the earliest stages of group assembly, and to explore their correlations with the mass and fractional abundance of this component. Additionally, we evaluate the potential of the IGL as a luminous tracer of the total mass distribution in galaxy groups, with particular focus on systems that are far from being dynamically relaxed.}
   {We use controlled numerical simulations of 100 low-mass galaxy groups spanning a range of masses and numbers of constituent galaxies to track the formation and evolution of IGL during the earliest pre-virialization stages of these systems.}
   {We show that the IGL typically begins to form in significant amounts after the turnaround epoch, which in our simulated groups occurs at a median redshift $\bar{z}_{\rm{ta}}\sim 0.85$, increasing steadily thereafter. We observe a sublinear relationship between the masses of this component and the brightest group galaxy, indicating intertwined formation histories but differing growth rates, which suggests that other group members may also significantly contribute to the diffuse light. Additionally, we observe indications that IGL formation is enhanced in groups with lower internal velocity dispersions, suggesting that gravitational interactions among member galaxies become more effective when their relative velocities are reduced. Two-thirds of our groups reveal significant alignment between the radial surface density profiles of the total and IGL mass, with fractional discrepancies below $25\%$. This supports the notion that this diffuse and faint baryonic component serves as a reliable tracer of the total gravitational potential in galaxy aggregations, regardless of their dynamical state. However, the results also indicate that the degree of similarity depends on the viewing direction.} 
   {}


\titlerunning{Growth of Ex-situ Diffuse Intragroup Light}
\authorrunning{B. Bilata-Woldeyes et al.}
\maketitle
%
\section{Introduction}

The diffuse intragroup light (IGL) is a pervasive feature of galaxy groups\footnote{Throughout the paper, we will use the word 'group' indistinctly to refer to both galaxy groups and the more densely populated galaxy clusters.} consisting of an extended low-surface-brightness component that permeates the intergalactic medium of these galaxy associations. It is primarily formed by stars that have been incorporated into the intragroup medium after being separated from their host galaxies \citep{2006ApJ...648..936R, 2009ApJ...699.1518R, 2019A&A...622A.183J}, although a portion may also be created in situ during the disruptive gravitational interactions experienced by group members \citep{2022ApJ...930...25B, 2023MNRAS.521..800M, 2024OJAp....7E.111A}, mainly during the formation stage of these galaxy systems \citep{2003ApJ...589..752G, 2010MNRAS.406..936P,2016MNRAS.461..321S}. Often found concentrated around the central, brightest group galaxy \citetext{BGG; \citealt{2015IAUGA..2247903M, 2021Galax...9...60C, 2022NatAs...6..308M}}, the IGL is a valuable source of information regarding the group's built-up and evolutionary stage \citep{2006ApJ...648..936R}. Additionally, several investigations suggest that this component of faint and diffuse light may also trace the dark matter distribution within these galaxy systems independently of their dynamical state \citep{2019MNRAS.482.2838M, 2020ApJ...901..128C, 2022ApJS..261...28Y}.

The significant variations in the outcomes of IGL research largely arise from the absence of a universally accepted definition for this diffuse component. As a result, its measurement is shaped by the differing approaches used in observational and simulation-based methodologies to identify and quantify its contribution to the total light of the groups \citep{2011ApJ...732...48R, 2012MNRAS.425.2058B, 2014MNRAS.437.3787C, 2016ApJ...820...49J, 2018MNRAS.479..932C, 2021A&A...651A..39R, 2022FrASS...952810R, 2023A&A...670L..20R, 2022NatAs...6..308M, 2023Natur.613...37J, 2023MNRAS.518.1195M, 2024MNRAS.528..771B}. While certain metrics achievable in simulations cannot be directly replicated in observations, complicating direct comparisons, the primary observational challenges stem from the proximity in the projected phase space between the IGL and the largest group galaxies. This proximity makes it difficult to define clear boundaries and precisely distinguish between these two components \citep{2011ApJ...732...48R, 2021ApJ...910...45M, 2022MNRAS.514.3082M}.

Among the various methods proposed in the literature to quantify the IGL fraction -- defined as the ratio of IGL to total light --, the application of an upper limit on the surface brightness (SB) to separate galactic from intergalactic starlight is by far the most extensively used technique \citep{2005MNRAS.358..949Z, 2011ApJ...732...48R, 2014MNRAS.437..816C, 2018ApJ...859...85T, 2018ApJ...862...95K, 2018MNRAS.474..917M, 2022NatAs...6..308M, 2022ApJ...925..103C}. Simulations have shown that this approach is effective for SB values fainter than $\mu_V\gtrsim 26.0$--$26.5$ mag arcsec$^{-2}$, approximately corresponding to the Holmberg radius, concurrently revealing that most of the IGL fraction was produced in a relatively recent cosmic epoch \citetext{$z\lesssim 1$; e.g. \citealt{2011ApJ...732...48R, 2014MNRAS.437.3787C, 2015ApJ...809L..21M, 2019A&A...622A.183J, 2021MNRAS.502.2419F, 2022NatAs...6..308M, 2022ApJ...925..103C}}. Contrastingly, other studies have proposed alternative interpretations
\citep[e.g.][]{2018MNRAS.474..917M,2023Natur.613...37J}. 

However, comparing these results with observations remains challenging due to discrepancies in photometric bands, aperture sizes, image depths, and SB cutoff thresholds, as noted by \citet{2011ApJ...732...48R} and \citet{2022NatAs...6..308M}. Among these factors, the lack of a universally accepted SB cutoff value is arguably the most significant limitation affecting IGL studies using this method, whether observational or simulation-based \citep[see also][]{2014A&A...565A.126P, 2015MNRAS.449.2353B, 2017MNRAS.467.4501H, 2021ApJS..252...27K}.

Another widely used strategy for detecting the IGL involves fitting its projected radial distribution with a combination of two or more functional forms, such as a S\'ersic profile and an exponential \citep[e.g.][]{2021ApJS..252...27K, 2021A&A...651A..39R, 2022FrASS...952810R, 2023A&A...670L..20R}. This approach is based on the assumption that the extended halo of the BGG and the IGL follow different projected surface brightness distributions. By modelling these components separately, the technique produces a composite radial light profile with distinct segments, allowing for an effective separation of the IGL from the BGG \citep[e.g.][]{2015ApJ...809L..21M, 2022NatAs...6..308M}. A dynamic variation of this method incorporates spectroscopic data to measure the velocity dispersion of starlight along the line of sight as a function of radial distance from the system's centre \citep{2011ApJ...732...48R, 2011MNRAS.414..602T, 2015ApJ...809L..21M, 2015A&A...579A.135L, 2022A&A...663A..12H}. The objective is to identify the radius at which the projected velocity dispersion of the BGG begins to rise to eventually converge with the velocity dispersion profile of the host galaxy group. This technique may prove effective in massive, dynamically relaxed groups with a well-defined central, first-ranked galaxy. However, it has very limited effectiveness in low-mass galaxy systems, where the velocity dispersions of the IGL and BGG are too similar to be unambiguously distinguished.

In this paper, we build on the widespread use of the SB method for detecting IGLto examine how various physical parameters of galaxy groups influence the ex-situ formation of this diffuse component during the earliest stages of group assembly. We also explore the potential of using the IGL as a luminous tracer of the overall gravitational potential of galaxy groups, particularly in systems that are far from dynamic equilibrium. To address these objectives, we performed simulations of moderately massive galaxy groups with total masses ranging from around $1$ to $5\,\times\,10^{13}\,M_\odot$\footnote{These systems represent the most common galaxy associations, encompassing at least $50\%$ of the galaxies in the local volume.}. Our simulations are designed to trace the early formation stages of these groups, focusing on the production of ex-situ IGL prior to their full dynamical relaxation (Section~\ref{sec:data}). In Section~\ref{sec:properties}, we outline the group properties employed in this study. The results are presented in Section~\ref{sec:result}, beginning with the BGG-based classification of the groups (Section~\ref{sec:BGGprop}), followed by an assessment of the likelihood that specific group properties align with their BGG class (Section~\ref{sec:grpprop}). We then examine the growth of both the fractional abundance and total mass of ex-situ IGL produced during the initial gravitational collapse of the groups, prior to their virialization, and compare our findings with values reported in the literature for galaxy systems spanning a variety of masses and dynamical states (Section~\ref{sec:IGLprop}). In Section~\ref{sec:correlations}, we explore correlations between IGL-related parameters and various group properties. Finally, in Section~\ref{sec:profiles}, we discuss the agreement between the projected IGL and total mass density distributions in our simulated groups. We conclude with a summary of our findings and final remarks in  Section~\ref{sec:conclusions}

In the present investigation, cosmology-dependent physical quantities have been calculated assuming a standard flat Lambda Cold Dark Matter ($\Lambda$CDM) cosmology with $H_0 = 70\;\mbox{km}\;\mbox{s}^{-1}\;\mbox{Mpc}^{-1}$, $\Omega_{\mathrm{m,0}}=0.3$, and $\Omega_{\Lambda,0}=0.7$.
\begin{figure*}[!ht]
\centering
\includegraphics[width=18.5cm, height=10cm]{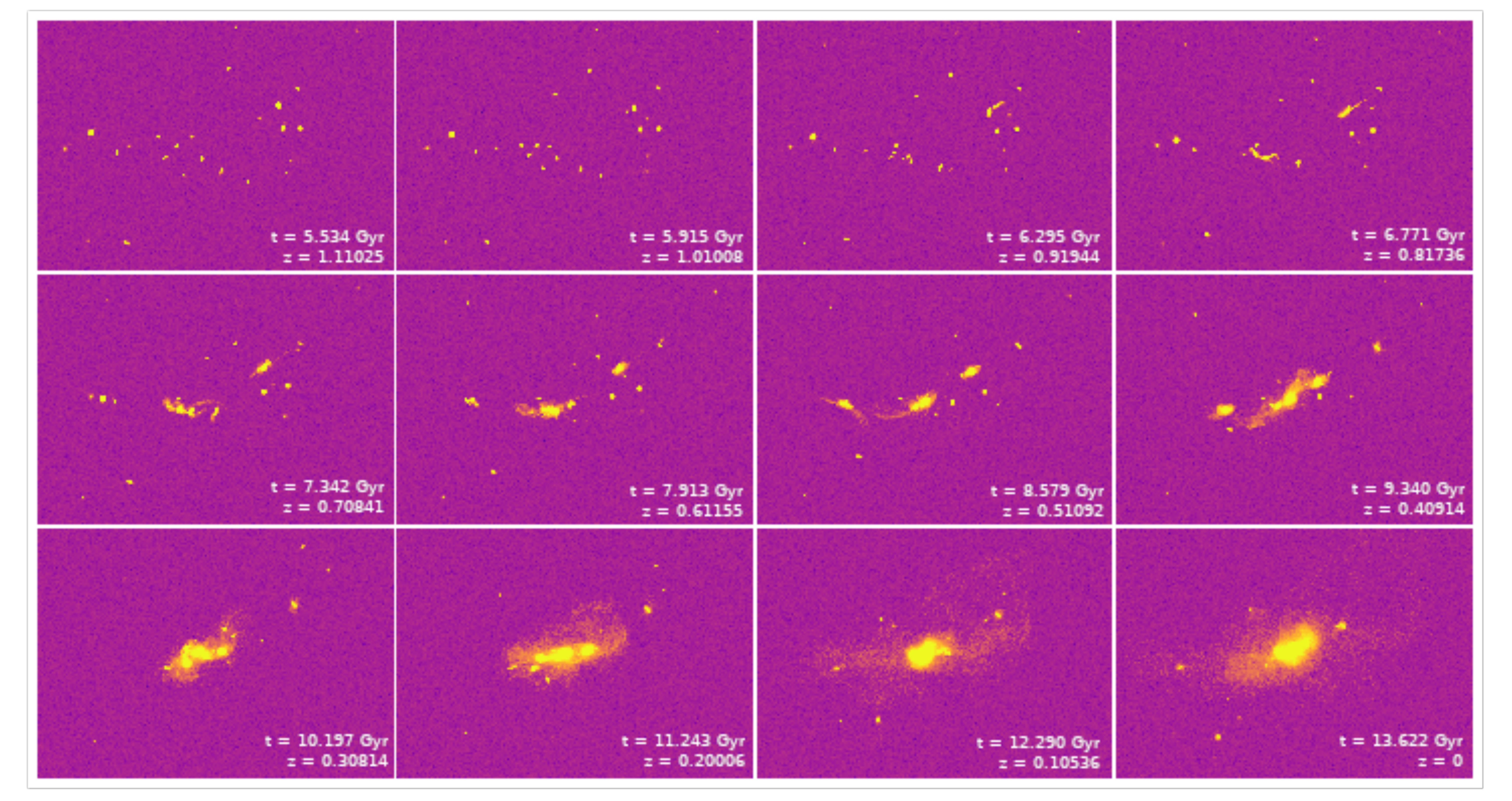}
    \caption{Twelve snapshots showing the evolution of one of our groups along $\sim 8$~Ga, from the turnaround at $z \simeq 1.1$ (upper-left panel) to $z = 0$ (lower-right panel).} 
    \label{fig:1951}
\end{figure*} 

\section{Numerical runs} \label{sec:data}
\subsection{Simulating pre-virialized galaxy groups} \label{ssec:simulations}
Our analysis of the IGL is based on measurements extracted from a batch of 100 controlled galaxy group simulations, evolved using the N-body run type of the cosmological code GADGET-2 \citep{2005MNRAS.364.1105S}.

The groups are created as nearly uniform isolated spherical overdensities that first expand linearly, then turnaround, and finally undergo a completely nonlinear collapse. Assuming that pressure gradients are negligible, each galaxy group is represented by a spherical top-hat perturbation of amplitude $\delta_{\rm i} > 0$ at the cosmic time $t_{\rm i}$ corresponding to the initial redshift $z_{\rm i}=3$ adopted for the simulations. This overdensity evolves like a Friedmann universe of initial radius 
\begin{equation}\label{overdensity}
R_{\mathrm{grp}}(t_{\rm i})=\left[\frac{3 M_{\mathrm{grp}}}{4\pi\rho_{\rm crit}(t_{\rm i})\Omega_{\mathrm m}(t_{\rm i})(1+\delta_{\rm i})}\right]^{1/3}\;,
\end{equation}
where the critical density, $\rho_{\rm crit}(t_{\rm i})$, and the mass density parameter, $\Omega_{\rm m}(t_{\rm i})$, refer to the unperturbed background cosmology. The value of the initial perturbation representing the groups, $\delta_{\rm i}$, is chosen so that a perfectly homogeneous overdensity with the same total mass of the group, $M_{\rm{grp}}$, would collapse into a single point at the final redshift $z_{\mathrm f} = 0$ of the runs. With this setup, the initial phase of the simulation, spanning roughly the first third of the total run time ($\sim 11.5$~Ga in the adopted cosmology\footnote{Throughout this paper, we use the symbol 'Ga' to denote 'billions of years,' in accordance with the conventions established by the International Astronomical Union (see https://www.iau.org/publications/proceedings\_rules/units/).}), is devoted to allowing the galaxies sufficient time to reach complete dynamical equilibrium within the group halo and to develop realistic peculiar velocities and $n$-point spatial correlations. This preparatory stage ensures that our simulated galaxy groups provide an accurate representation of these systems prior to the onset of widespread mergers during the subsequent nonlinear gravitational collapse stage that follows the turnaround epoch.

The model galaxies consist of a rotating extended spherical cold dark matter (CDM) halo with a Navarro–Frenk–White density profile \citep{1997ApJ...490..493N}, whose global properties (mass, spin, and concentration) are used to scale the central baryonic (stellar) core. The luminous component's mass is set at $5\%$ of the total mass \citep[e.g.][]{Governato+2007, Sales+2009, Dai+2010}, distributed either in an exponential disc plus an inner spheroidal bulge following the \citet{1990ApJ...356..359H}'s profile, or in a single \citeauthor{1990ApJ...356..359H} spheroid. We maintain an equal split in number between luminous and dark bodies across all galaxy models. The Plummer equivalent softening length for luminous particles is fixed at 30 pc, sufficiently small to accurately resolve the dynamics of any simulated galaxy, as well as other structures of interest, such as tidal streams and shells. For the more massive CDM bodies, the softening length is adjusted proportionally to the square root of their body mass, thereby ensuring consistent maximum interparticle gravitational forces across the simulations.
\begin{figure*}[!ht]
\centering
      \includegraphics[width=16.2cm, height=8.4cm]{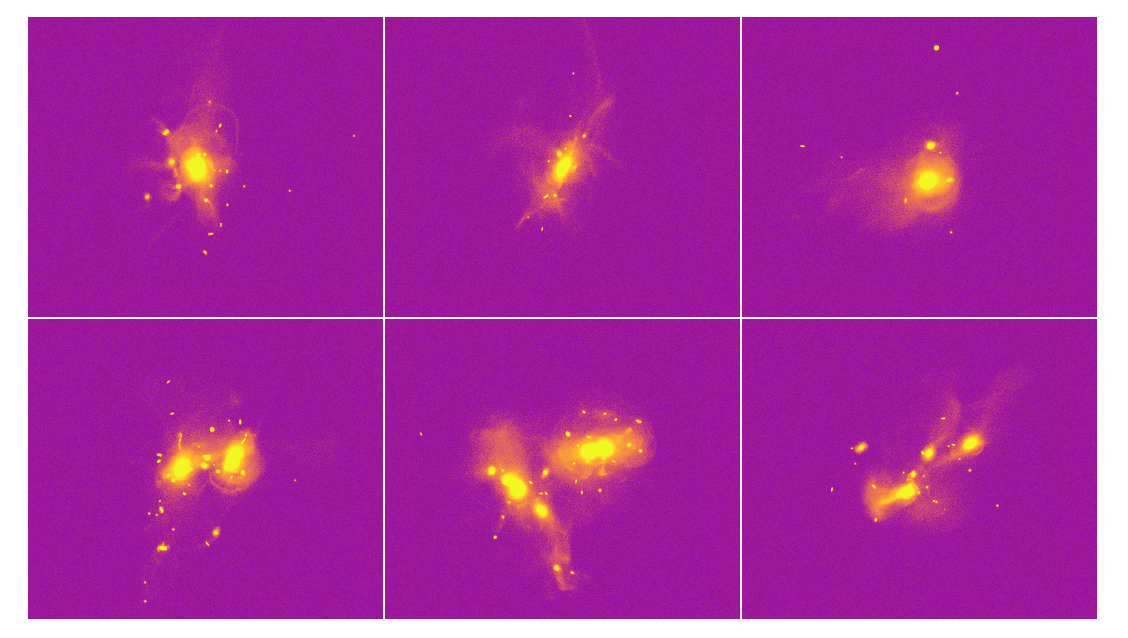}
    \caption{Examples of group images at $z=0$. Upper panels: groups with a clear dominant galaxy or BGG (see text). Lower panels: groups with two dominant galaxies (left) or lacking a truly dominant BGG (middle and right). Note that in all cases, the IGL is mainly distributed around the largest galaxies.}
    \label{fig:mix}
\end{figure*}
The total (virial) masses of the galaxy haloes initially contained in the simulated groups are randomly drawn from a Schechter probability distribution function (PDF) with an asymptotic slope of $\alpha=-1.0$ and a characteristic mass of $M^*=10^{12}\,h^{-1}\,\mbox{M}_\odot$. We impose a lower mass limit on this PDF of $M_{\rm min}=0.05\,M^*$ since smaller haloes do not significantly impact the results. All galaxies with masses below $0.1\,M^*$ are assumed, for simplicity, to possess purely spheroidal stellar distributions\footnote{We have verified that the structural and kinematic properties of galaxies above this mass limit remain statistically unchanged for as long as a Hubble time when they are evolved in isolation.}. Beyond this mass threshold, galaxy morphologies are determined using a Monte Carlo technique that assumes an initial late-type galaxy fraction of $0.80$ on average, consistent with the morphological distribution of the JWST galaxies \citep{Lee+2024}. The initial structural and dynamical properties of all central stellar distributions are adjusted to align with the main scaling laws defined by real objects of the same morphology in the local universe. Comprehensive details on the modelling of the baryonic cores are provided in  \citet{2018A&A...614A..66S}, as well as in \citet{2010A&A...516A...7D}.

After establishing the central distributions of the progenitor galaxies, we have adjusted the total masses and taper radius of their CDM haloes by scaling them down to values consistent with the initial redshift of the simulations. However, we have left the stellar cores unchanged, as most mergers in our experiments occur at relatively late epochs when galaxy properties closely resemble those observed in the Local Universe. Furthermore, while our simulations do not explicitly model the earlier formation phase of the galaxies' progenitors, it is important to note that the structure and dynamics of remnants from dry major mergers are predominantly shaped by the orbital parameters of the collisions, rather than the intrinsic properties of the merging objects, owing to the high energies involved in such events. Once all galaxy haloes within a given group have been generated and rescaled, the common uniform background of the group is filled evenly with CDM particles identical to those constituting individual galaxies, until the total group mass is achieved. The initial positions of background particles and the centre of mass of member galaxies are randomly distributed within the group volume, ensuring avoidance of overlapping galaxy haloes, while their initial velocities follow the local Hubble flow. All multi-component galaxy models employed in the experiments achieve complete dynamical equilibrium with the surrounding group halo during the expansion phase of the groups, well before they reach the turnaround epoch (see also Section~\ref{sec:IGLprop}).

In establishing the initial number of galaxies per group, $N_{\rm gal}(z_{\rm i})$, we have implemented a criterion ensuring that the final tally of galaxies in our simulations is consistent with the statistics of subhalo abundance in typical group-sized $\Lambda$CDM haloes \citep{2011MNRAS.410.2309G}. This is accomplished by treating $N_{\rm gal}(z_{\rm i})$ as a random Poisson variable with parameter $\lambda$ (i.e.\ mean and variance) given by the empirical formula
\begin{equation}\label{N_memb}
\lambda\equiv N_{\rm bright}(z_{\rm i})\frac{M_{\rm grp}}{10M^{\rm su}}\frac{\Phi(\alpha,M_{\rm min})}{\Phi(\alpha,0.5M^*)}\;,
\end{equation}
with $N_{\rm bright}(z_{\rm i})$ representing the initial number of massive (luminous) galaxies (i.e.\ with $M_{\rm gal}\geq 0.5\;M^*$) and $\Phi(\alpha,x)$ the cumulative Schechter mass function (MF) integrated up to $x$, and where $M^{\rm su}$ is the unit mass in our simulations, which we take equal to the characteristic mass adopted for the MF of galaxies\footnote{$M^*$ is also the total halo mass commonly associated with a MW-sized galaxy \citep[see, e.g.][]{2010MNRAS.406..896B}.}. In the current series of simulations, we opt for $N_{\rm bright}(z_{\rm i})=5$. This choice, alongside the selected parameter values, yields a spectrum of $N_{\rm gal}(z_{\rm i})$ within our groups that scales linearly with the total mass of these systems, spanning from 20 to 69 and with a median value settling at 31\footnote{Since the initial radius of the overdensity representing our smallest galaxy aggregations is near 1 Mpc, this initialization process implies that all our groups initially contain about five bright galaxies per Mpc$^3$.}. The first, second (median), and third quartiles $(Q_1,Q_2,Q_3)$ of the distribution of the total mass of our galaxy systems, $M_{\rm grp}$, are, respectively, $(1.63,2.13,2.76)\,\times10^{13}\,\mbox{M}_{\odot}$, while the values of these same dividers for the distribution of the total stellar mass, $M_\star$, are $(6.87, 8.43, 11.2)\,\times\,10^{11}\,\mbox{M}_{\odot}$. The total-mass-to-stellar-mass-ratio of the groups ranges between $15$ and $43$, with a median value of $24$. Additional information regarding the methodology employed to configure our model groups and galaxies can be found elsewhere \citep{2016MNRAS.461..321S,2016MNRAS.461..344P,2018A&A...614A..66S}. 

\subsection{Mapping the IGL and identifying galaxies} \label{ssec:identification}
To map the IGL and study its evolution, we identified and isolated this diffuse component from the member galaxies of the groups in three independent (i.e., orthogonal) projections of each of the 120 snapshots -- evenly spaced in cosmic time from $z_{\mathrm{i}} = 3$ to $z_{\mathrm{f}} = 0$ -- into which we divided the runs.

We began by determining the distribution of the IGL at different epochs, isolating the less dense regions of our groups through a SB cutoff in the simulated images. Following \citet{2005ApJ...631L..41M}, \citet{2011ApJ...732...48R}, and \citet{2014MNRAS.437..816C}, among others, we applied a projected stellar density threshold equivalent to a SB of $26.5$ $V$ mag arcsec$^{-2}$ to identify the IGL in our simulations.
This threshold was first converted into physical units of solar luminosity per square parsec using the well-established conversion formula
\begin{equation}\label{muV}
\mu_{V}(\mbox{mag}\,\mbox{arcsec}^{-2})={\cal{M}_{V,\odot}} + 21.572 - 2.5\log\mu(L_{V,\odot}\,\mbox{pc}^{-2})\;,     
\end{equation}
with ${\cal{M}}_{V,\odot}=4.81$ mag the absolute visual magnitude of the Sun. The resulting value was then converted into a stellar surface density by translating luminosity into stellar mass. This was done by adopting a uniform stellar mass-to-$V$-band-light ratio of $5\,\Upsilon_{V,\odot}$, the same value used by \citet{2011ApJ...732...48R} in their simulated galaxy clusters to account for the older stellar populations characteristic of galaxy associations. This yields a 2D stellar mass density threshold of $4.5\,\mbox{M}_{\odot}\,\mbox{pc}^{-2}$, which we applied to all stellar particles across snapshots from our 100 group runs to trace the evolution of the IGL. As \citeauthor{2011ApJ...732...48R} emphasize, one implication of taking a constant density threshold to identify the IGL at all cosmic times is that we are excluding the effects of stellar evolution from the dynamical evolution of this component. The density estimation is obtained by means of the nearest neighbour particle estimation in both 2D and 3D versions. The selected KDtree algorithm was the efficient Fortran-tree software implemented by Li Dong\footnote{https://github.com/dongli/fortran-kdtree} and in our case, we searched until the $50^{th}$ nearest particle.

Once IGL particles were identified in our simulations, we separated them from the rest of the light in each one of the group images and used the remaining particles to determine, with the aid of the centroid-based Mean-Shift unsupervised algorithm \citep[see, e.g.][]{Carreira2015}, the 3D position of the centroids of individual galaxies. This algorithm operates on the principle of moving each particle towards the mode, or peak density, within a specified radius, an iterative process that continues until the points converge to a local maximum of the density function. The candidate local maxima are then filtered in a post-processing stage to eliminate near-duplicates to form the final set of centroids, therefore obviating the need to predict the number of galaxies in each snapshot beforehand. The number of galaxies in any given snapshot corresponds to the count of maxima identified at the end of this procedure. Both the kernel bandwidth parameter, which determines the window size for galaxy identification, and the galaxy centroids, were estimated using only the luminous particles located in the densest regions of the snapshots. This minimizes the problems that can potentially affect the identification of galaxies due to the strong overlap that their outer regions may register in advanced mergers. After some experimentation with different density thresholds in the first half (chronologically speaking) of our experiments, we found that this galaxy finding algorithm performed optimally when applied to luminous particles with 3D local densities exceeding the 65th percentile of their ranked distribution of densities in a snapshot. We have verified that with this setup, the positions and number of galaxies retrieved with the Mean-Shift clustering algorithm are closely aligned with those produced by the \textsc{SExtractor} software \citep{1996A&AS..117..393B} in any of the images that can be constructed from the three orthogonal projections of the snapshots. 

After determining the centroids of all galaxies in the group images, we redistributed the luminous particles that were not classified as IGL among the galaxies to determine their stellar masses. To assign non-IGL particles to individual objects, we applied a recursive procedure based on the linking strength, $f_i$, between each particle and each centroid, defined by the empirical relation:
\begin{equation}\label{force}
f_i\equiv\frac{m_i}{r_i^4}\;,     
\end{equation}
where $m_i$ represents the stellar mass of galaxy $i$ and $r_i$ is the 3D distance of the particle from the galaxy's centroid\footnote{The fourth power in $r$ is chosen to ensure fast convergence of the procedure.}. In the first step, $m_i$ was replaced by the luminous mass of the galaxy's densest inner region, defined, in accordance with the method used to identify galaxy centroids, as the mass within the region surrounding the galaxy that encompasses the top $35\%$ of luminous particles with the highest local densities. In the second step, the remaining non-IGL particles were allocated to the galaxy whose centroid yielded the highest value of $f_i$. After distributing the $65\%$ of non-IGL particles with the lowest local densities among the galaxies, the linking strengths were recalculated. At this stage, the $m_i$ represents the total stellar mass of the galaxy $i$, determined by summing the masses of all assigned non-IGL particles, which we subsequently ranked in descending order. This second step was repeated iteratively until convergence was achieved. Note that projecting the 3D spatial data from the snapshots onto the Cartesian planes effectively triples the sample size for the majority of the parameters under investigation (see Sec.~\ref{sec:properties}). This approach also offers rudimentary yet valuable insights into the variance within the simulated data, arising from factors such as object identification, potential contamination from nearby companions, and projection effects\footnote{The number of particles associated with the galaxies and IGL can vary in a non-monotonic way from one snapshot to the next as a result of the stochastic tidal interactions and mergers that take place during the gravitational collapse of the groups.}. 

The attentive reader will have noticed a potential inconsistency between our definitions of IGL and the strategy we have just followed to determine the total luminous mass of galaxies. We have defined the IGL using a 2D surface density threshold to facilitate direct comparisons with observational data, which is our primary focus. In contrast, defining the IGL as the 3D stellar material not associated with galaxies would introduce variability in the projected maximum density of this component, complicating comparisons between simulations and observations. Preliminary tests have confirmed that this inconsistency has a negligible impact on our results\footnote{The typical difference between the total mass in stellar particles included in each simulation and the luminous mass we measure from combining the IGL particles (identified using a 2D surface density threshold) with those associated with galaxies (identified using a 3D linking length) is around $1\%$.}.

\section{Selected group properties} 
\label{sec:properties}
We have measured a series of parameters from all images generated by projecting the snapshots in which we divided our group simulations onto the three Cartesian planes defined by the coordinate axes. These properties include $M_{\rm grp}$, $M_\star$, the total mass-to-luminous mass ratio, $M_{\rm grp}/M_\star$, the total mass of the IGL, $M_{\rm IGL}$, the IGL fraction, $f_{\rm IGL}\equiv M_{\rm IGL}/M_{\rm \star}$, as well as the mass-weighted velocity dispersion of the group from the equation
\begin{equation}\label{barV}
\sigma_{\rm{grp}}= \sqrt{\frac{1}{M_{\rm gal}} \sum_{i=1}^{N_{\rm gal}} m_{i}(v_{i} -\bar{v})^2}\;,   
\end{equation}
where $v_{i}$ is the line-of-sight (LOS) component of the velocity of the $i$-th galaxy, $\bar{v}$ the mass-weighted velocity of the centroid of the galaxy distribution
\begin{equation}\label{vmean}
{\bar{v}}= \frac{1}{M_{\rm gal}} \sum_{i=1}^{N_{\rm gal}} m_{i}v_{i}\;,   
\end{equation}
and $M_{\rm{gal}}=\sum_i m_i$ is the total luminous mass in galaxies. We also infer a representative scale radius for each galaxy group image in the form of the mass-weighted observed mean separation between  galaxies given by the equation
\begin{equation}\label{Rgrp}
\bar{R}_{\rm{gal}}=\frac{2}{{M^2_{\rm gal}}}\sum_{i<j}^{N_{\rm gal}-1} \sum_{j=i+1}^{N_{\rm gal}} {m_{i}m_{j}}{r_{ij}}\;,     
\end{equation}
with $r_{ij}$, the projected separation between galaxy pairs. Table~\ref{tab:props} lists the key summary statistics for the distributions of the four group properties at $z = 0$ that exhibit variations depending on the line of sight, measured across all three orthogonal projections.
\begin{table}
     \caption[]{LOS-dependent group properties measured at $z=0$.}
	\label{tab:props}
\resizebox{\columnwidth}{!}{
	\begin{tabular}{l|c|c|c}
		\toprule
    \textbf{Property} & $\bm{XY\,\mbox{plane}^*}$ & $\bm{XZ\,\mbox{plane}^*}$ & $\bm{YZ\,\mbox{plane}^*}$ \\
    		\midrule
           $\sigma_{\rm grp}$ (km s$^{-1}$) & $155^{+33}_{-50}$  &   $145^{+46}_{-35}$  &  $149^{+48}_{-40}$ 
           \\
           & & & \\
           $\bar{R}_{\rm gal}$ (kpc) & $294^{+169}_{-68}$  &   $305^{+165}_{-76}$  &  $332^{+141}_{-98}$       \\
           & & & \\
           ${f}_{\rm IGL}$ $(\%)$ & $11.4^{+1.5}_{-1.7}$  &   $11.1^{+2.1}_{-1.5}$  &  $11.3^{+1.6}_{-1.6}$     \\
           & & & \\
           ${M}_{\rm IGL}$ $(10^{10}\;M_\odot)$ & $9.83^{+2.87}_{-2.75}$  &   $9.78^{+3.82}_{-2.68}$  &  $9.90^{+3.6}_{-2.71}$       \\
        \bottomrule
        \noalign{\smallskip}
     \end{tabular}}
     \small 
     (*) Orthogonal projections; quoted values are medians and interquartile ranges.
\end{table}

Finally, we have introduced two parameters that identify the different types of first-ranked galaxies hosted by our groups. These parameters compare the luminous masses of the three largest group members, expressed in the familiar form of magnitude differences between the first and second-ranked galaxy\footnote{First-ranked galaxies ($m_1$) are the most massive galaxies in the group in terms of stellar mass. Those with a magnitude gap with respect to the second most massive object ($m_2$) above a certain threshold are considered true BGGs (see Sec.~\ref{sec:BGGprop}).}, $\Delta{\cal{M}}_{\rm{2-1}}$, and between the second and third-ranked galaxy $\Delta{\cal{M}}_{\rm{3-2}}$. We calculate these differences by converting the corresponding galaxy masses into $K$-band magnitudes, assuming a stellar mass-to-light ratio of one $\Upsilon_{K,\odot}$.

\section{Results} \label{sec:result}
We open this section with a summary of the ex-situ growth of the IGL in our simulations, juxtaposing our findings with prior observational and theoretical studies on the abundance of this component. A comprehensive analysis of the cosmic evolution of both the fractional abundance and total mass of the IGL during the early formation phase of low-mass galaxy groups will be presented in a forthcoming paper (Bilata-Woldeyes et al., in preparation).
\subsection{Growth of IGL in the earliest stages of group assembly} \label{sec:IGLprop}
\begin{figure}[!ht]
	\includegraphics[width=8.5cm, height=5cm]{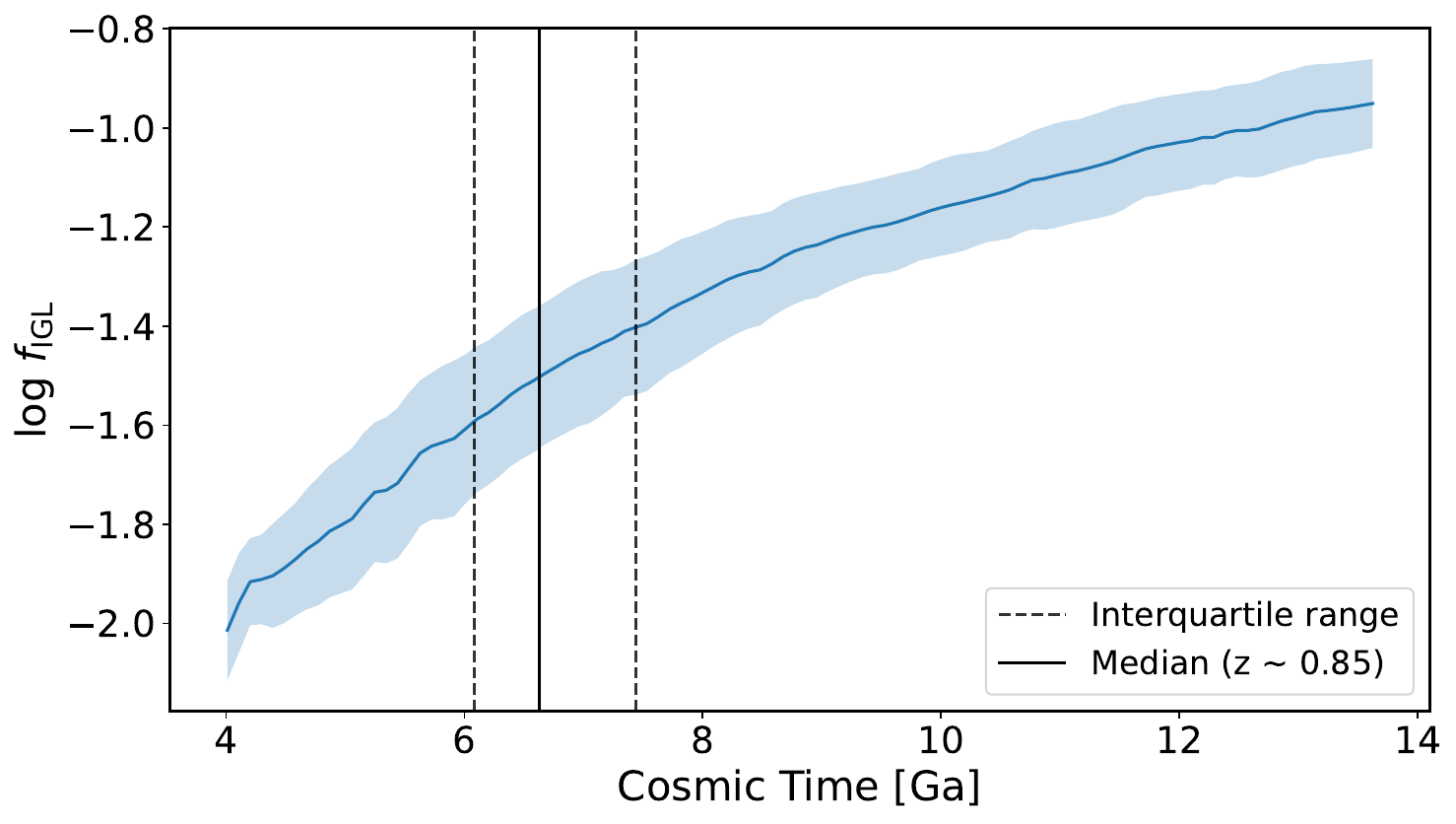}
    \caption{Growth of the IGL fraction over cosmic time in our simulated groups. The blue band represents the interquartile range of the temporal variation in the distribution of $\log(f_{\rm{IGL}})$ in the last $\sim\!10$~Ga of the runs, with the central solid line indicating the evolution of the median value. The vertical solid and dashed lines mark, respectively, the $Q_2$, and $Q_1$ and $Q_3$ quartiles of the distribution of turnaround times, when the groups reach their largest non-expanding scale around the centre of gravity.} 
    \label{fig:AllMod_fIGL_Gyr}
\end{figure}
\begin{figure}
\centering
	\includegraphics[width=8.5cm, height=5.5cm]{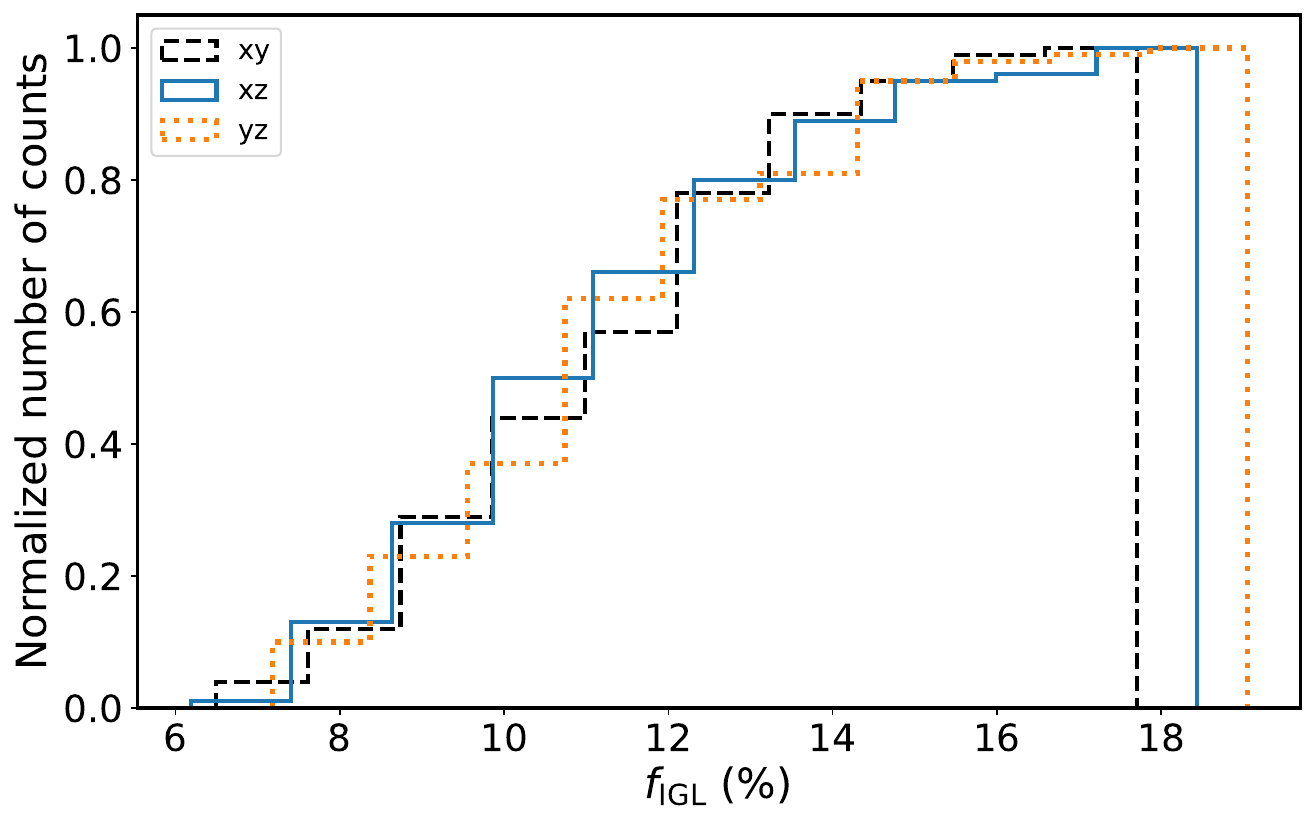}
    \caption{Cumulative distribution of the ex-situ IGL fraction in our group sample at $z = 0$, measured across all three orthogonal projections. IGL particles along the lines of sight passing through galaxies are excluded, resulting in a slight underestimation of the IGL fraction and mass which introduces a mild variation in the assessment of this component from different viewing angles. See Table~\ref{tab:props} for the location and scale parameters of the corresponding PDFs.}
    \label{fig:fIGL_distribution}
\end{figure}    

As shown in Fig.~\ref{fig:AllMod_fIGL_Gyr}, $f_{\rm IGL}$ begins to grow $\sim 2$~Ga after the start of the simulations. Notably, while the initial relative growth of IGL is faster and approximately linear, the majority of this diffuse and faint component forms after the groups reach their turnaround epoch (at a median redshift $\bar{z}_{\rm{ta}}\sim 0.85$ in our simulations), aligning with findings from previous studies  \citep{2007MNRAS.377....2M, 2011ApJ...732...48R, 2015MNRAS.449.2353B, 2016MNRAS.461..321S}. Beyond this point, the ongoing hierarchical collapse of these systems shortens intergalactic distances, increasing the frequency and intensity of gravitational interactions and enhancing the efficiency of stellar stripping among member galaxies \citep[e.g.][]{2015MNRAS.448.1162D, 2018ApJ...857...79J, 2018MNRAS.474.3009D, 2019A&A...622A.183J}. The quartiles of the distribution of the total ex-situ IGL mass generated in our runs at $z = 0$, measured along the three orthogonal projections, are $(Q_1,Q_2,Q_3)=(7,10,13)\,\times 10^{10}\,\mbox{M}_{\odot}$, while the corresponding IGL fractions range between $\sim 6$--$18\%$, with a median value of $11.3\%$ (see Fig.~\ref{fig:fIGL_distribution} and Table~\ref{tab:props}). These ex-situ fractions can be contrasted with the total diffuse light fractions reported in previous both observational and theoretical studies. For example, while \citet{2014ApJ...791...38W} found no detectable IGL in the low-mass Leo I group, \citet{2020A&A...642A..46H} reported a low limit for the IGL fraction of approximately $4\%$, whereas \citet{2022FrASS...952810R} measured $17\%$, all using surface brightness-based estimates but with varying bands and depths. Likewise, \citet{2023A&A...671A..83G} measured a $f_{\mathrm{IGL}}\sim 20\%$ in a virialized group with a mass of $\sim 6 \times 10^{13}\,M_\odot$ detected in the foreground of the MACS0329 galaxy cluster.¡ Conversely, estimates of the diffuse light fraction in the Coma cluster by \citet{2019A&A...622A.183J} range from $\sim 7$--$21\%$, depending on the filter used, while for the Virgo cluster fractions between $\sim 7$ and $15\%$ were reported in the $V$ band \citep[e.g.][]{2017ApJ...834...16M}. In the local universe ($z \lesssim 0.05$), studies of IGL in multiple compact groups by \citet{2005MNRAS.364.1069D} and \citet{2008MNRAS.388.1433D} reported fractions ranging from 0 to $50\%$. Additionally, \citet{2023A&A...670L..20R} compiled data from a homogeneous sample of 22 galaxy groups and clusters, finding $f_{\mathrm{IGL}} \sim 5$--$45\%$, with a best-fit value of approximately $18\%$ within the virial mass range covered by our simulations. At higher redshifts ($z\sim 0.2$–$0.5$), the fractions of intracluster light tend to fall between $\sim 5$ and $35\%$, varying with the cluster dynamics and filters applied \citetext{e.g. \citealt{2021MNRAS.501.1300S, 2023MNRAS.518.1195M}}. For their part, $N$-body simulations of galaxy clusters that apply the SB cutoff method with thresholds similar to ours have reported intergalactic light fractions around $10$–$15\%$ in evolved systems \citep{2006ApJ...648..936R, 2011ApJ...732...48R}. Taking into account the significant differences in the detection methodologies and field extensions, as well as in the masses and dynamical histories of the galaxy systems investigated, we conclude that the results of our simulations are reasonably consistent with the fractional abundances of diffuse light reported across these studies.

\subsection{BGG-based classification of groups} \label{sec:BGGprop}
We now turn our attention to the classification of groups based on the presence or absence of a dominant brightest group galaxy (BGG). This classification is informed by visual inspection of group images, which reveals that each realization of the initial conditions in our group model gives rise to a diverse range of assembly paths. These evolutionary paths leave distinct imprints on the number and relative sizes of the group galaxies, particularly the top-ranked members, at the end of the runs. 

Using the magnitude gaps described in Sec.~\ref{sec:properties}, we define a subset of groups, termed single-BGG groups, as those that at $z=0$ host a clearly dominant galaxy that is at least $0.75$ magnitudes brighter than the next brightest member\footnote{To our knowledge, the literature lacks a standardized definition for the minimum magnitude gap that identifies BGGs. Our chosen threshold reflects established conventions and is optimized to produce group subclasses with distinctly different behaviours in the correlations examined.}. This single-BGG class comprises $38\%$ of the group images at the end of the simulations, with a significant fraction ($14\%$ of all images) exhibiting extreme magnitude gaps $\Delta{\cal{M}}_{2-1}\geq 1.7$ mag, characteristic of fossil groups \citep{2016ApJ...824..140R}. On the other hand, $16\%$ of our systems are classified as double-BGG groups, defined by the conditions $\Delta{\cal{M}}_{2-1} < 0.5$ mag and $\Delta{\cal{M}}_{3-2} \ge 0.75$ mag, to indicate the presence of a pair of bright galaxies that stand out from the rest of group members. The remaining $46\%$ of the $z=0$ images correspond to groups termed non-BGG since they have both $\Delta{\cal{M}}_{2-1}$ and $\Delta{\cal{M}}_{3-2} < 0.75$ mag. These latter systems, often the most massive (see Sec.~\ref{sec:iglvsprops}), typically exhibit multiple ongoing mergers but lack truly dominant objects. See also Table~\ref{tab:BGGs} and the examples of each group class depicted in Fig~\ref{fig:mix}.
\begin{table}
\centering
     \caption[]{Group classes according to BGG content.}
	\label{tab:BGGs}
	\begin{tabular}{c|c|c}
		\toprule
        \textbf{Group class} & \textbf{Definition} & \textbf{Fraction}   \\ 
    		\midrule
        Single-BGG & $\Delta{\cal{M}}_{2-1} \geq 0.75$ mag & 0.38 \\[0.2cm]
        Double-BGG & {\renewcommand{\arraystretch}{1.3}\begin{tabular}{c@{}} $\Delta{\cal{M}}_{2-1} < 0.5$ mag \\ $\Delta{\cal{M}}_{3-2} \ge 0.75$ mag \end{tabular}} & 0.16 \\[0.5cm]
        Non-BGG & {\renewcommand{\arraystretch}{1.3}\begin{tabular}{c@{}} $\Delta{\cal{M}}_{2-1} < 0.75$ mag \\ $\Delta{\cal{M}}_{3-2} < 0.75$ mag \end{tabular}}  & 0.46 \\
    		\bottomrule
           	\end{tabular}
\end{table}
\begin{figure}
\centering
      \includegraphics[width=0.8\columnwidth, height=5.55cm]{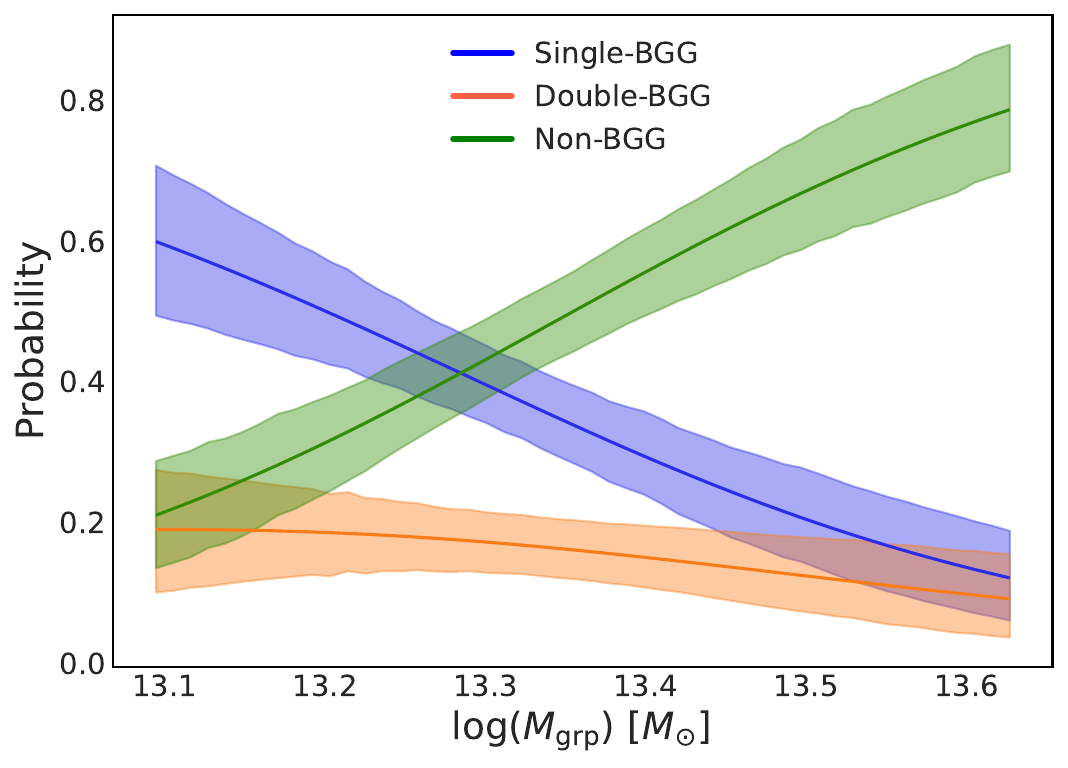} 
      \includegraphics[width=0.8\columnwidth, height=5.55cm]{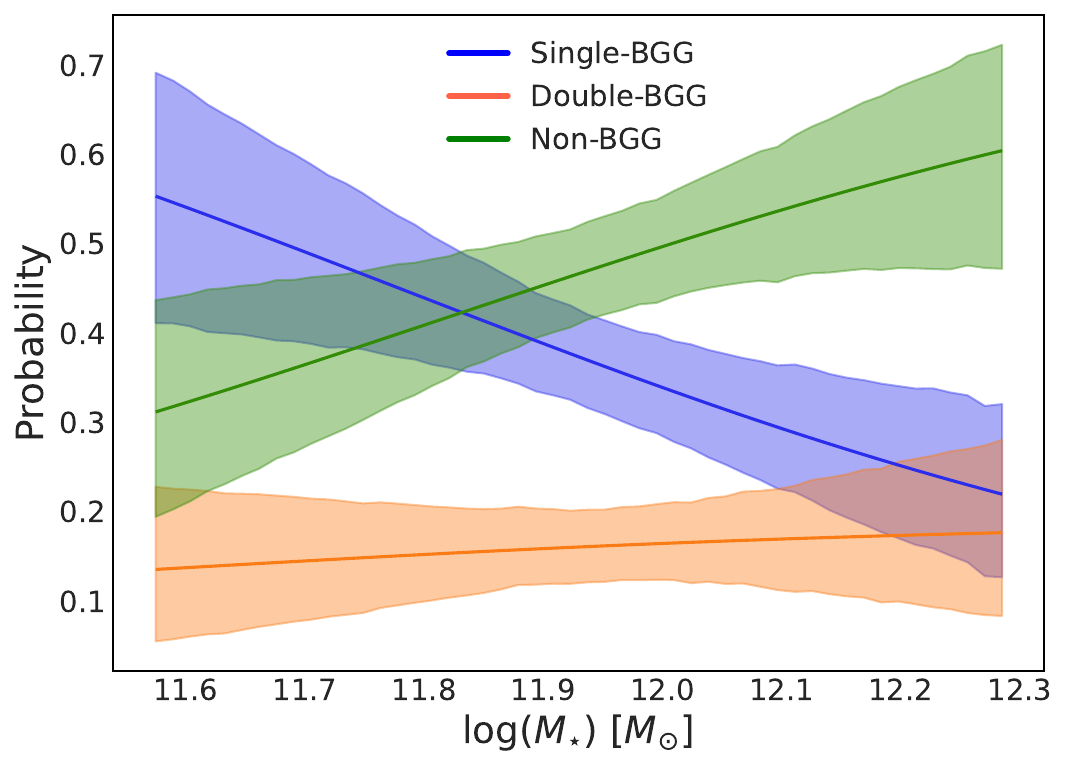} 
      \includegraphics[width=0.8\columnwidth, height=5.55cm]{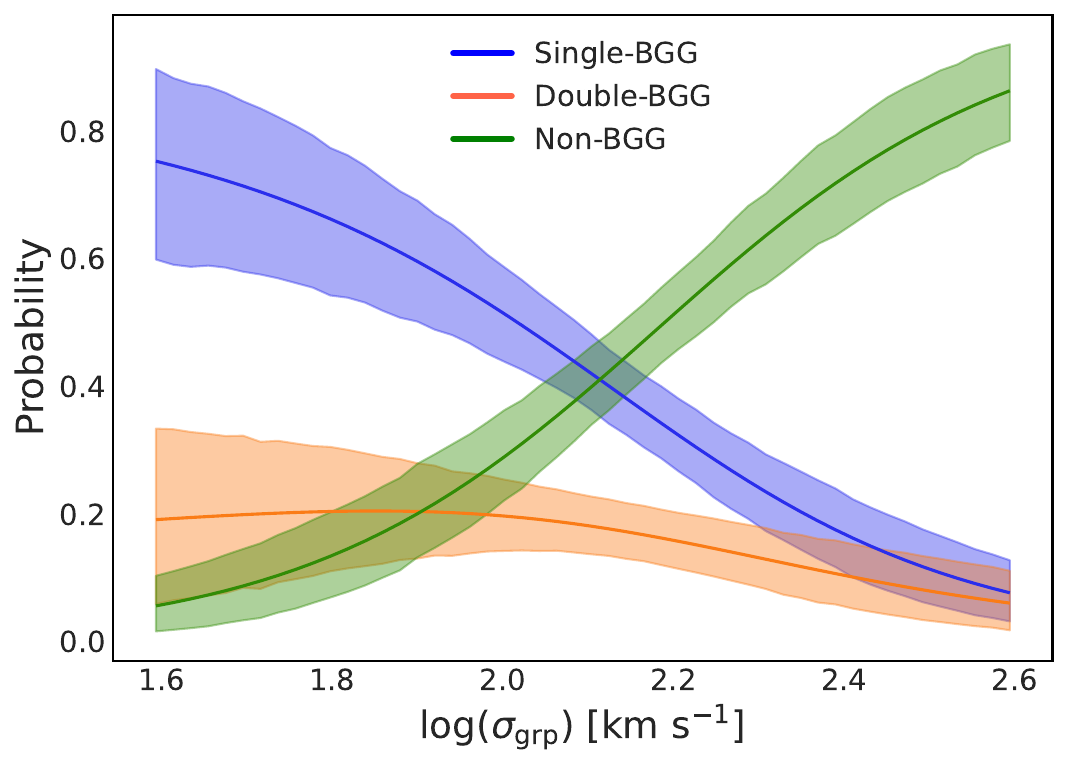} 
      \includegraphics[width=0.8\columnwidth, height=5.55cm]{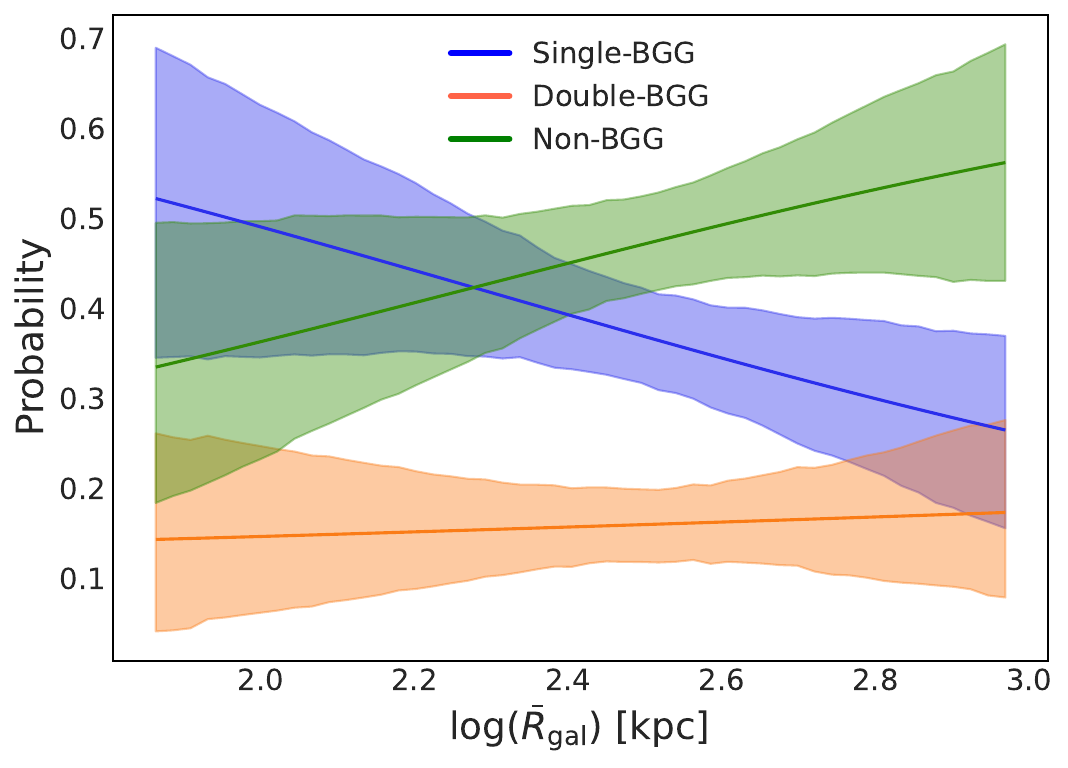} 
    \caption{Probability that a given group parameter takes on a certain value by BGG class. From top to bottom: total system mass, total stellar mass, velocity dispersion, and mean projected intergalactic separation. Shaded bands show interquartile ranges for single-BGG (blue), double-BGG (orange), and non-BGG (green) groups, with solid lines indicating medians.}
    \label{fig:dominant}
\end{figure}
\begin{figure*}
\centering
 	\includegraphics[width=5.9cm, height=4.6cm]{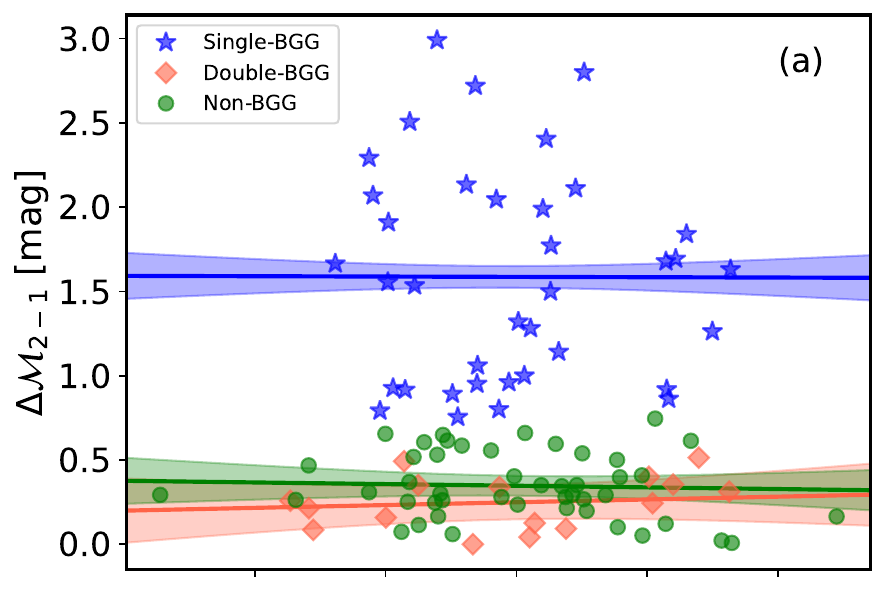}
    \includegraphics[width=5.3cm, height=4.6cm]{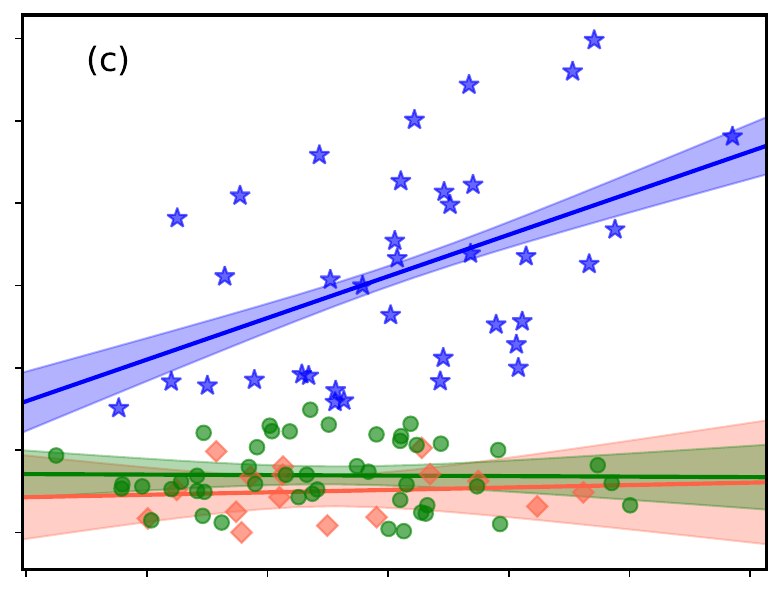}
    \includegraphics[width=7cm, height=4.6cm]{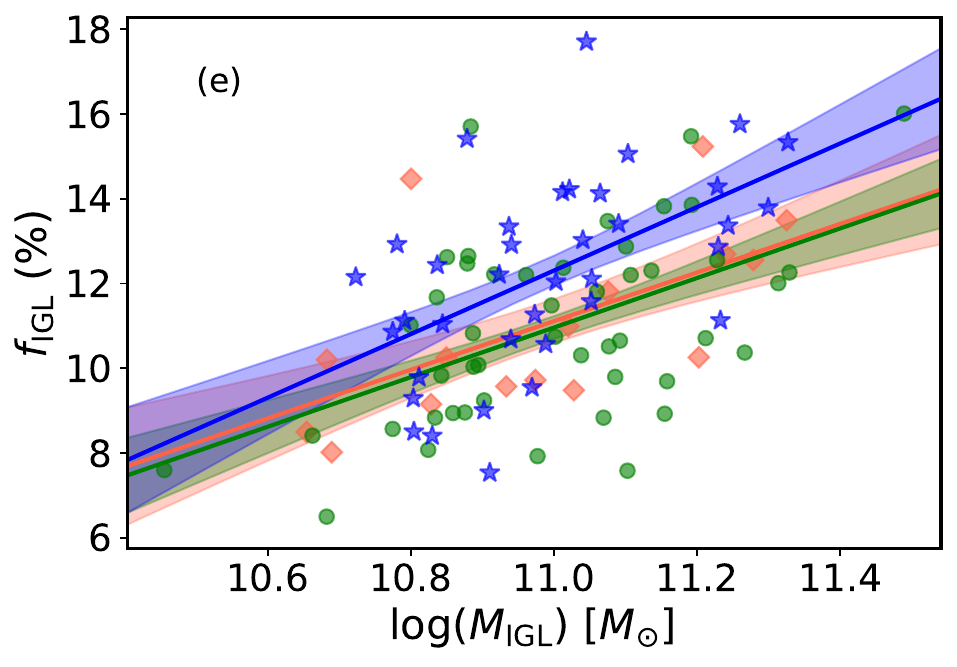}    
    \includegraphics[width=5.9cm, height=5.1cm]{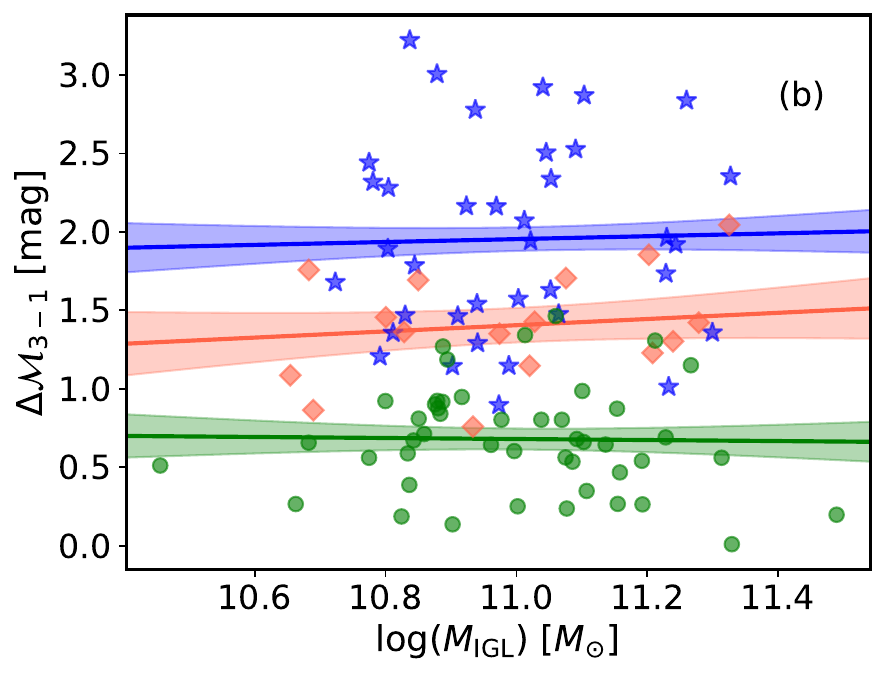}
    \includegraphics[width=5.3cm, height=5.1cm]{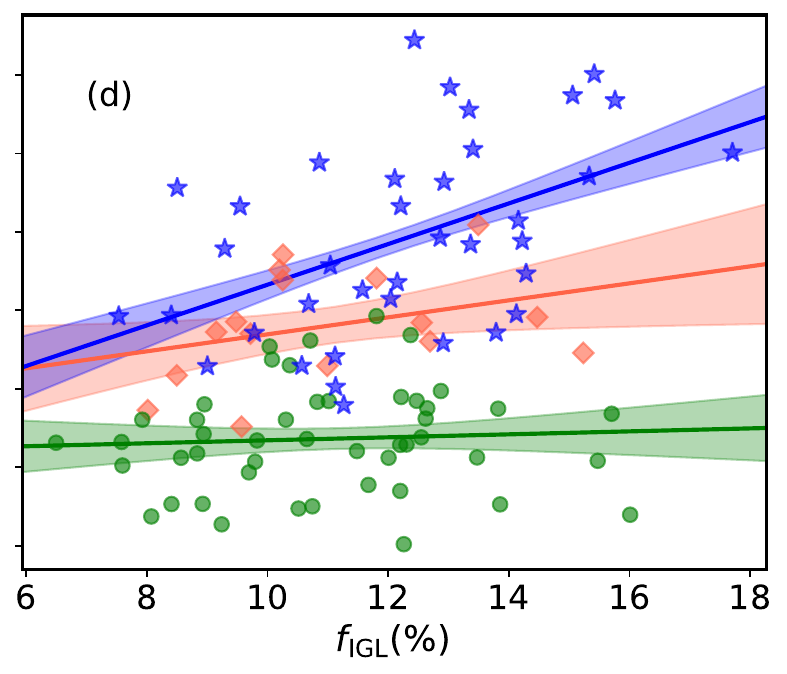}
    \includegraphics[width=7cm, height=5.1cm]{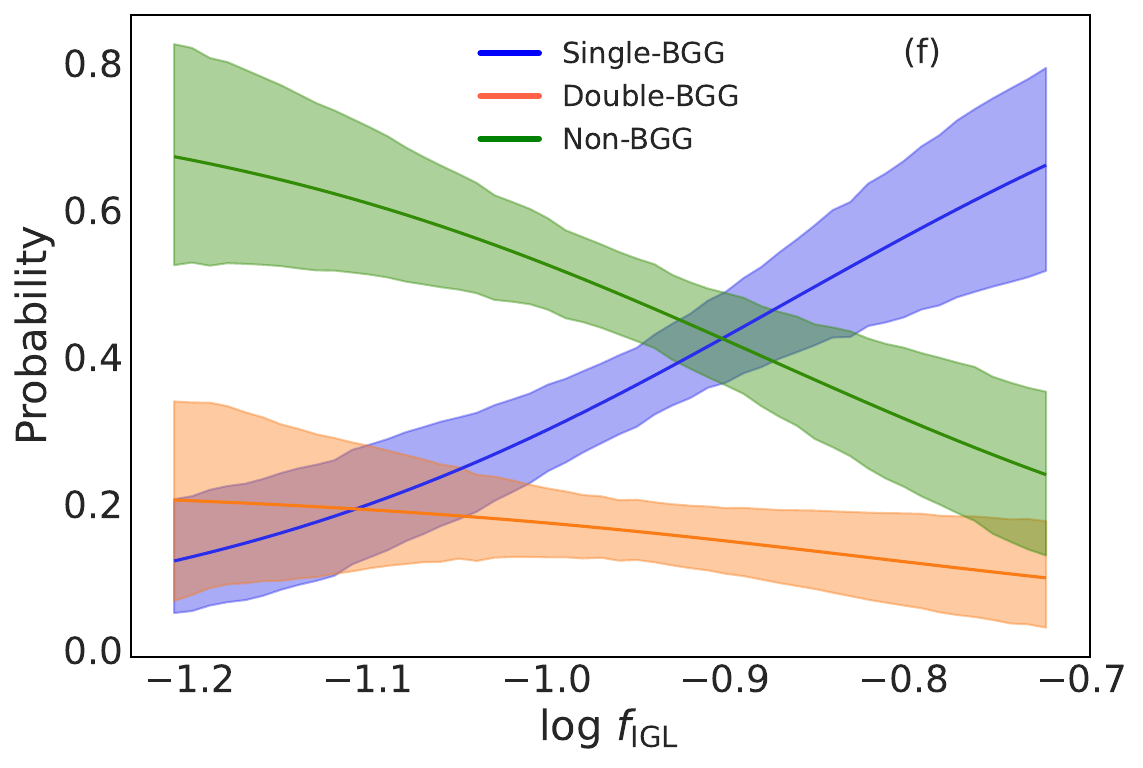}
    \caption{(a)--(b) Logarithm of the IGL mass and (c)--(d) IGL fraction plotted, respectively, against the $\Delta{\cal{M}}_{2-1}$ and $\Delta{\cal{M}}_{3-1}$ magnitude gaps of the BGG. (e) The IGL fraction vs the logarithm of the IGL mass. In panels (a)--(e) blue stars represent single-BGG groups, for which $\Delta{\cal{M}}_{2-1} \geq 0.75$ mag, orange diamonds denote double-BGG groups, with $\Delta{\cal{M}}_{2-1} \leq 0.5$ mag and $\Delta{\cal{M}}_{3-2} \geq 0.75$ mag, while green circles correspond to non-BGG groups, where both $\Delta{\cal{M}}_{2-1}$ and $\Delta{\cal{M}}_{3-2}$ are less than $0.75$ mag indicating the absence of a clearly dominant BGG. The solid lines and shaded bands in these panels illustrate, following the same colour scheme, linear fits and their associated uncertainties for each BGG subset. The fits are included to highlight the moderately strong positive linear correlation (Pearson's $r=0.46$ and $r=0.49$, respectively) that exists between $f_{\mathrm{IGL}}$ and $\Delta{\cal{M}}_{2-1}$ and $\Delta{\cal{M}}_{3-1}$ for the single-BGG group subset (panels c and d), as well as the mild positive relationship ($r = 0.33$) observed between $f_{\mathrm{IGL}}$ and $\Delta{\cal{M}}_{3-1}$ for the double-BGG group subset (panel d). Since the magnitude gaps are view-independent, data in all these panels are limited to single-axis projections of the $z=0$ snapshots for clarity, showing only measurements in the $XY$ plane. (f) Same as Fig.~\ref{fig:dominant}, but for the IGL fraction at $z=0$. In this case, the probabilities are derived from the 300 measurements of $f_{\rm IGL}$ obtained using the three orthogonal projections.}
    \label{fig:mgap_mIGL_figure}
\end{figure*}

\subsection{Group properties according to BGG class} 
\label{sec:grpprop}
We next show the relative probabilities and associated uncertainties that a given group property takes on a certain value according to the BGG class of the group. The properties depicted in the four panels of Fig.~\ref{fig:dominant} are, from top to bottom, (the logarithms of the) total system mass, total stellar mass, group velocity dispersion, and mean intergalactic separation. The colour bands showing the fractional abundances of the group classes have been calculated using a multicategory logistic model, known as the baseline-category logit, that employs all three BGG classes at once by defining the odds of the outcome in one class compared to the others \citep{agresti2012categorical} as implemented in the Python package Bambi \citep{Capretto2022}. The response probability for a given category is obtained using the general formula
\begin{equation}\label{pi}
\pi_{\rm j}= \frac{\exp{(\beta_{\rm j0}+\beta_{\rm j1}x_{\rm 1}+\beta_{\rm j2}x_{\rm 2}+...+\beta_{\rm jp}x_{\rm p})}}{\sum_{h=1}^{c} \exp{(\beta_{\rm h0}+\beta_{\rm h1}x_{\rm 1}+\beta_{\rm h2}x_{\rm 2}+...+\beta_{\rm hp}x_{\rm p})}}\;, \ \ j = 1,\ldots,c    
\end{equation}
where $c$ denotes the total number of categories or classes ($c = 3$ in our case), $x_i$ are the $p$ values of one of the explanatory variables mentioned above that sample its distribution, and $\beta_{ji}$ the coefficients of the multinomial logistic regression model. The sum of the responsive probabilities of all the categories must be one, i.e.\ $\sum_{\rm j} \pi_{\rm j} = 1$.

The main result of this comparison is the contrasting behaviour between single-BGG groups and non-BGG systems across all properties examined within the ranges covered by our simulations. Single-BGG groups show the highest fractional abundance at the lower end of the distributions for total and stellar group mass, group velocity dispersion, and projected intergalactic separation. Conversely, the highest values of all these parameters are the heritage of non-BGG systems. Double-BGG systems, on the other hand, exhibit a much more neutral behaviour, with fractional abundances showing minimal variations: decreasing slightly with increasing total mass and velocity dispersion and raising marginally with higher stellar mass and intergalactic separation.

At this point, however, it is important to emphasize that the presence or absence of a distinct BGG does not appear to significantly influence either the total amount of diffuse light produced or its resulting fraction. This is best illustrated in panels (a)--(d) of Fig.~\ref{fig:mgap_mIGL_figure}, which show the $\Delta{\cal{M}}_{2-1}$ and $\Delta{\cal{M}}_{3-1}$ magnitude gaps as a function of the total mass and fraction of this component for the three BGG group classes. As these graphs demonstrate, all BGG group types exhibit a comparable range of $\log(M_{\mathrm{IGL}})$, although the most extreme values are found in non-BGG groups. Similarly, the range of $f_{\mathrm{IGL}}$ values is fairly consistent across the three categories, although single-BGG groups show, on average, slightly higher fractions of diffuse light compared to the other two classes. Single-BGG groups also exhibit moderately strong positive linear correlations between $f_{\mathrm{IGL}}$ and $\Delta{\cal{M}}_{2-1}$ and $\Delta{\cal{M}}_{3-1}$ magnitude gaps, with Pearson's $r=0.46$ and $r=0.49$, respectively. In double-BGG groups, however, this positive relationship ($r=0.33$) is observed only with respect to $\Delta{\cal{M}}_{3-1}$. By contrast, in non-BGG groups, the IGL fraction shows no significant correlation with either magnitude gap, just as with happens with the null dependence of the total IGL mass and these gaps across all group types. On the other hand, panel (e) of Fig.~\ref{fig:mgap_mIGL_figure} shows that the moderately strong positive linear correlation that exists between the fraction and mass of IGL produced by the groups is largely independent of their BGG classification ($r = 0.52$ for the entire group sample). Additionally, the response probability of the logarithm of the IGL fraction, depicted in panel (f), further underscores the contrasting behaviours of single-BGG and non-BGG group classes. Single-BGG groups dominate at the upper limit of $f_{\mathrm{IGL}}$ values, while non-BGG groups are more prevalent at the lower end. Consistent with previous findings, the relative abundance of double-BGG systems shows, by contrast, minimal variation, with only a slight decline as $f_{\mathrm{IGL}}$ increases.

\begin{figure*}[!ht]
\centering
      \includegraphics[width=15cm, height=10cm]{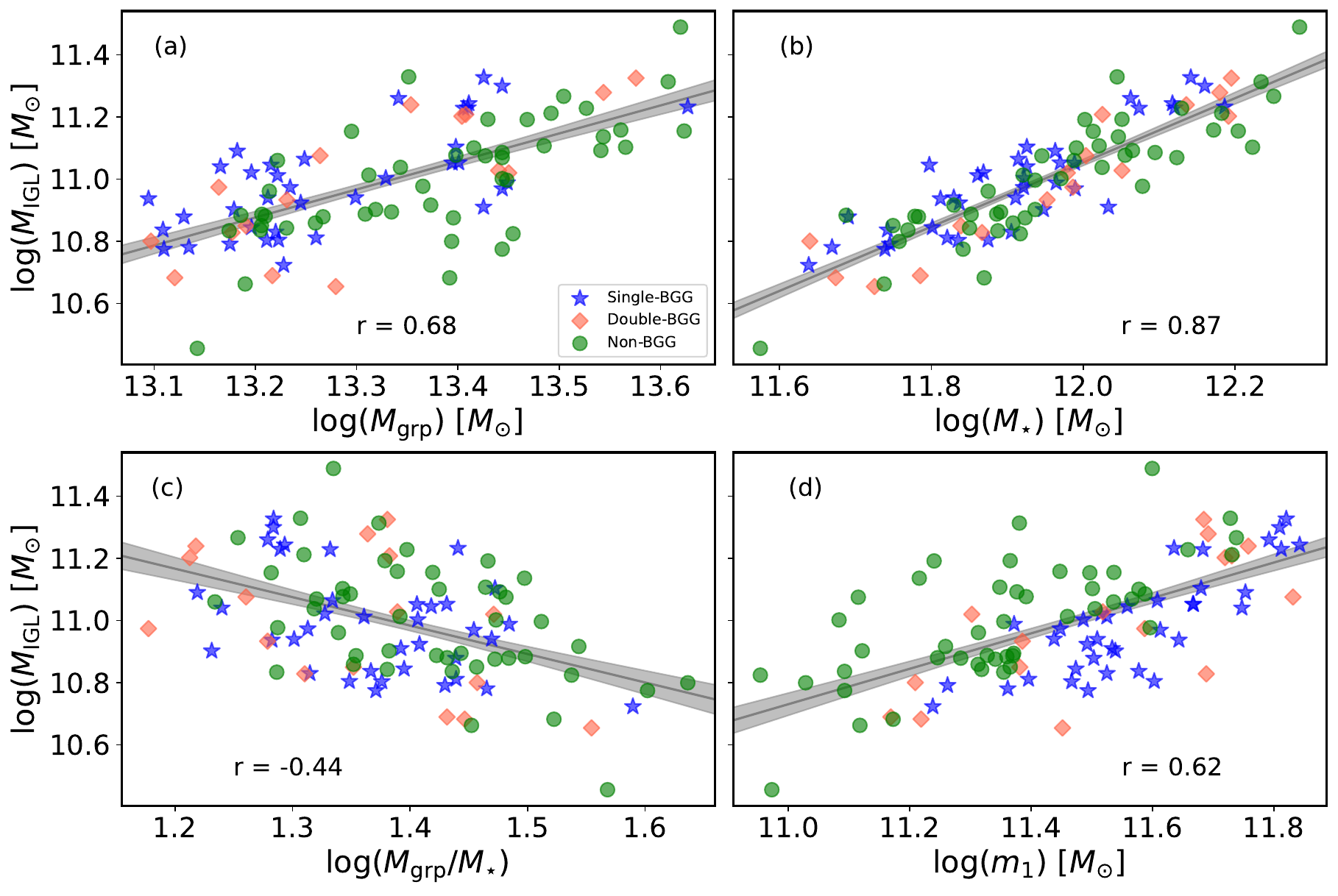} \hspace{0.3cm} \\
  \caption{Scatter plots of $M_{\mathrm{IGL}}$ against the four group properties with which this parameter is more strongly correlated, all measured at $z = 0$ (see Table~\ref{tab:statistics}). These properties are: (a) total system mass ($M_{\rm grp}$), (b) total stellar mass ($M_\star$), (c) total mass-to-stellar mass ratio ($M_{\rm grp}/M_\star$), and (d) mass of the first-ranked galaxy ($m_{1}$). Symbols for data points correspond to those used in Fig.~\ref{fig:mgap_mIGL_figure}. Since the values of all these four properties are unaffected by the viewing direction, we only include, for clarity, the data points corresponding to the projections onto the $XY$ plane. The grey solid lines and shaded bands represent the linear relationships and their associated $1\sigma$ uncertainties calculated for each pair of parameters. The corresponding values of the Pearson's correlation coefficients $r$ are provided in each panel.} 
  \label{fig:Mvisib_mICL_bar_figure}
\end{figure*}

\subsection{Correlations between group properties at $z=0$} \label{sec:correlations}
In this section, we investigate the relationships between the various group properties defined in Section~\ref{sec:properties} and the mass and fraction of the IGL, as well as the stellar mass of the first-ranked group galaxy, at the end of the simulations. Our aim is to assess whether the early stages of group assembly lead to relationships that can be modelled using linear regressions on log-log plots, effectively testing whether the pairs of variables constituting the arguments consistently follow a power law, and therefore exhibit scale invariance over the relatively narrow ranges of values explored\footnote{Due to this limitation, a straight-line on a log–log plot is a necessary, but not sufficient, condition to warrant a power-law relationship at all scales.}. Table~\ref{tab:statistics} summarizes the outcome of the statistical tests conducted, showing that in all cases the $p$-values of observing a non-zero log-log linear correlation exceed $0.5$. 

\begin{table}
     \caption[]{Correlations$^*$ between group properties at $z=0$.} 
	\label{tab:statistics}
	\begin{tabular}{lcrrrrc}
		\toprule
    $\bm{X}$ & $\bm{Y}$ & \multicolumn{1}{c}{$\bm{a}$}  & \multicolumn{1}{c}{$\bm{b}$} & \multicolumn{1}{c}{$\bm{r}$}  & \multicolumn{1}{c}{$\bm{t}$}  &$\bm{p}$  \\ 
    		\midrule
            $M_{\rm grp}$     &   $M_{\rm IGL}$  &    0.92  &  $-1.22$   &  0.68  &  16.20   & 1.0    \\
            $M_\star$         &   $M_{\rm IGL}$  &    1.03  &  $-1.31$  &   0.87  &   30.16 &  1.0   \\
            $M_{\rm grp}/M_\star$   &   $M_{\rm IGL}$  &    $-0.87$ &  12.20  &   $-0.44$  &  8.40   & 1.0   \\
            $m_{\rm{1}}$     &   $M_{\rm IGL}$  &   0.55  &  4.72   &  0.62  &  13.50 &  1.0       \\
            $\sigma_{\rm grp}$  &   $M_{\rm IGL}$  &   0.10 &  10.77  &  0.10  &   1.79 &  0.96   \\
            $\bar{R}_{\rm{gal}}$  &   $M_{\rm IGL}$  &   0.15 &  10.61  &  0.17  &   3.06  &  0.99   \\
            $M_{\rm grp}$     &   $f_{\rm IGL}$  &    $0.006$ &  $-1.02$  &  $0.01$  &  0.15   &  0.56    \\
            $M_\star$         &   $f_{\rm IGL}$  &    0.03  &  $-1.28$   &  0.05  &  0.82  &  0.79    \\
            $M_{\rm grp}/M_\star$   &   $f_{\rm IGL}$  &    $-0.07$ &  $-0.86$  &   $-0.07$  &  1.15   & 0.87   \\
            $m_{\rm{1}}$    &   $f_{\rm IGL}$  &  0.13  &  $-2.41$ &  0.29  &  5.25  &  0.99    \\
            $\sigma_{\rm{grp}}$  &   $f_{\rm IGL}$  &   $-0.16$ &  $-0.60$  &  $-0.33$  &   6.10  &  1.0   \\
            $\bar{R}_{\rm{gal}}$  &   $f_{\rm IGL}$  &   $-0.09$ &  $-0.75$  &  $-0.18$  &   3.23  &  0.99   \\
            $M_{\rm grp}$     &   $m_{\rm{1}}$  &  0.24  &  8.30  &  0.16 &  1.58  &  0.94     \\
            $M_\star$         &   $m_{\rm{1}}$  & 0.74  &  2.59  &  0.56  &  6.62  &  1.0    \\
            $M_{\rm grp}/M_\star$   &   $m_{\rm{1}}$  &  $-1.57$ &  13.64  &   $-0.70$  &  9.66   & 1.0   \\
            $\sigma_{ \rm grp}$  &   $m_{\rm{1}}$  &   $-0.31$ &  12.14  &  $-0.28$  &   5.03  &  0.99   \\
            $\bar{R}_{\rm{gal}}$     &   $m_{\rm{1}}$  &   0.03  &  11.40  &  0.03 &  0.46 &  0.68  \\     
        \bottomrule
        \noalign{\smallskip}
        \multicolumn{7}{l}{\footnotesize (*) Correlations are of the form $\log(Y)=a\;\!\log(X)+b$.}
         \end{tabular}
            \small
            Notes: $X$ and $Y$ - group properties, $a$ – slope, $b$ – intercept, $r$ – Pearson's correlation coefficient, $t$ – Student's $t$ value, $p$ – probability that the logarithms of the two parameters are linearly correlated when the null hypothesis is true.
\end{table}
\begin{figure*}
\centering
	\includegraphics[width=15cm, height=6cm]{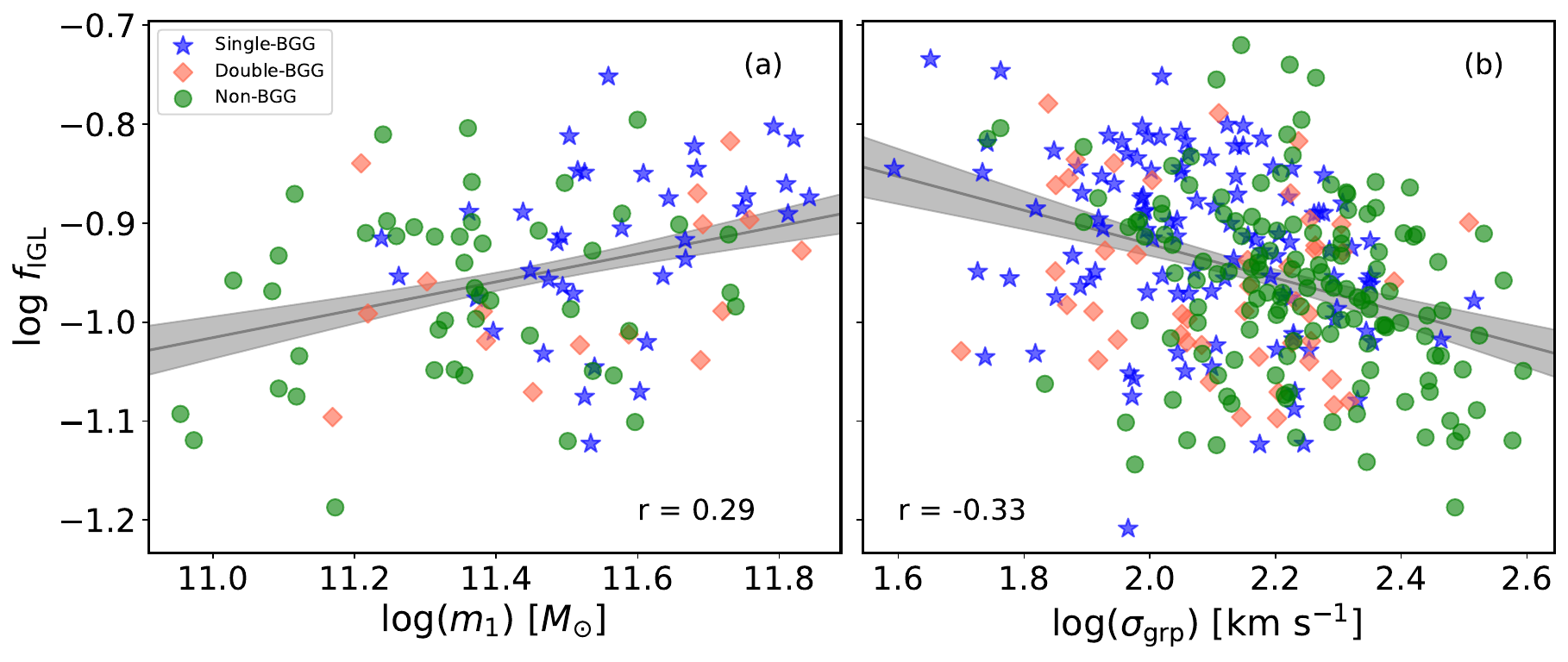}
     \caption{Same as Fig.~\ref{fig:Mvisib_mICL_bar_figure}, but for the correlation of $f_{\mathrm{IGL}}$ with (a) the mass of the first-ranked galaxy ($m_{1}$) and (b) the internal velocity dispersion of the groups ($\sigma_{\mathrm{grp}}$). Unlike the correlations depicted in panel (a) and in Figs.~\ref{fig:mgap_mIGL_figure} and \ref{fig:Mvisib_mICL_bar_figure}, each group in panel (b) is represented by three points, corresponding to three independent measurements on the Cartesian planes.} 
    \label{fig:vdisp_d5_linfit}
\end{figure*}

\subsubsection{Mass and fraction of IGL versus other group properties} \label{sec:iglvsprops}
As shown in Table~\ref{tab:statistics} (see also Fig.~\ref{fig:Mvisib_mICL_bar_figure}), $M_{\rm IGL}$ exhibits a very high likelihood of linear correlation ($p > 0.95$) with all the group parameters investigated when plotted on logarithmic axes. The strongest positive correlations of this parameter, ranked in descending order, are with the total stellar mass ($r=0.87$), total system mass ($r=0.68$), and the mass of the first-ranked galaxy ($r=0.62$). There is also a fourth relatively strong correlation involving the total mass-to-stellar mass ratio, which in this case is negative, reflected in a Pearson's $r$ of $-0.44$. The negative slope of this relationship indicates that groups with higher dark matter content (or, equivalently, lower stellar fraction) tend to produce smaller amounts of diffuse stellar light. Notably, three of these four log-log regression models align closely with the linear scaling of the arguments: positive for the independent variables $M_{\rm grp}$ and $M_\star$, and negative for $M_{\rm grp}/M_\star$. In contrast, we find that $m_1$ follows a positive sublinear relationship ($a=0.55$) with $M_{\rm IGL}$ (Fig.~\ref{fig:Mvisib_mICL_bar_figure}d), implying that the two components grow at different rates. While this result does not rule out the possibility that the formation of BGGs occurs in parallel with the build-up of diffuse stellar light, the non-linearity of the relationship suggests that galaxies other than the BGG may significantly contribute to the growth of diffuse light \citep[see, e.g.][]{2007MNRAS.377....2M, 10.1093/mnras/stv033}. This conclusion is further supported by the presence of substantial diffuse intergalactic light in some non-BGG groups. Additionally, Figs.~\ref{fig:Mvisib_mICL_bar_figure}a--b provide evidence that non-BGG groups are, on average, more massive and contain higher total stellar masses than their single and double-BGG counterparts.

In sharp contrast, the linear correlations involving $f_{\rm{IGL}}$ in logarithmic space are weak at best or, simply, non-existent. Among the former, the strongest are with $m_{\rm{1}}$ ($r=0.29$) and $\sigma_{\rm{grp}}$ ($r=-0.33$), as shown in Fig.~\ref{fig:vdisp_d5_linfit}. Additionally, we find that the linear trend in panel (a) is driven by the BGG-hosting groups (both single and double), which exhibit moderate correlations ($r = 0.32$ and $0.35$, respectively), whereas in non-BGG groups the IGL fraction is virtually uncorrelated ($r = 0.14$) with the mass of the first-ranked galaxy. On the other hand, the negative slope of the $f_{\rm{IGL}}$--$\sigma_{\rm{grp}}$ relation (Fig.~\ref{fig:vdisp_d5_linfit}b) hints, despite the weakness of the correlation, at an increased efficiency of gravitational forces in stripping material from galaxy discs when the interactions among group galaxies are slower \citep[see, e.g.][]{Mihos:2003, 2016ApJS..225...23S}. 

The remaining correlations in which $f_{\rm{IGL}}$ participates can be considered negligible, since all have $|r|\leq 0.20$. Notably, our simulations show a complete absence of any relationship between this variable and both the total group mass ($r=0.01$) and the total stellar mass ($r=0.05$), which is also reflected in the associated $p$-values, which are among the lowest inferred ($0.56$ and $0.79$, respectively). These results align with numerous previous studies \citep[e.g.][]{2021MNRAS.501.1300S, 2022NatAs...6..308M, 2023A&A...670L..20R, 2024A&A...683A..59C}, especially concerning the lack of correlation of $f_{\rm{IGL}}$ with total system mass \citep[e.g.][]{2010MNRAS.405.1544D, 2010MNRAS.403..768H, 2014MNRAS.437.3787C, 2015MNRAS.449.2353B, 2020MNRAS.494.1859A}. Nevertheless, there is no total consensus on the absence of such a connection, as some studies report a positive relationship between $f_{\rm{IGL}}$ and $M_{\rm grp}$ \citep[e.g.][]{2007MNRAS.377....2M, 2019MNRAS.489.2439H}, while others suggest that a mild negative correlation exists \citep[e.g.][]{2014MNRAS.437..816C}; see also the review by \citet{2021Galax...9...60C}.
\begin{figure*}
\centering
	\includegraphics[width=15cm, height=6cm]{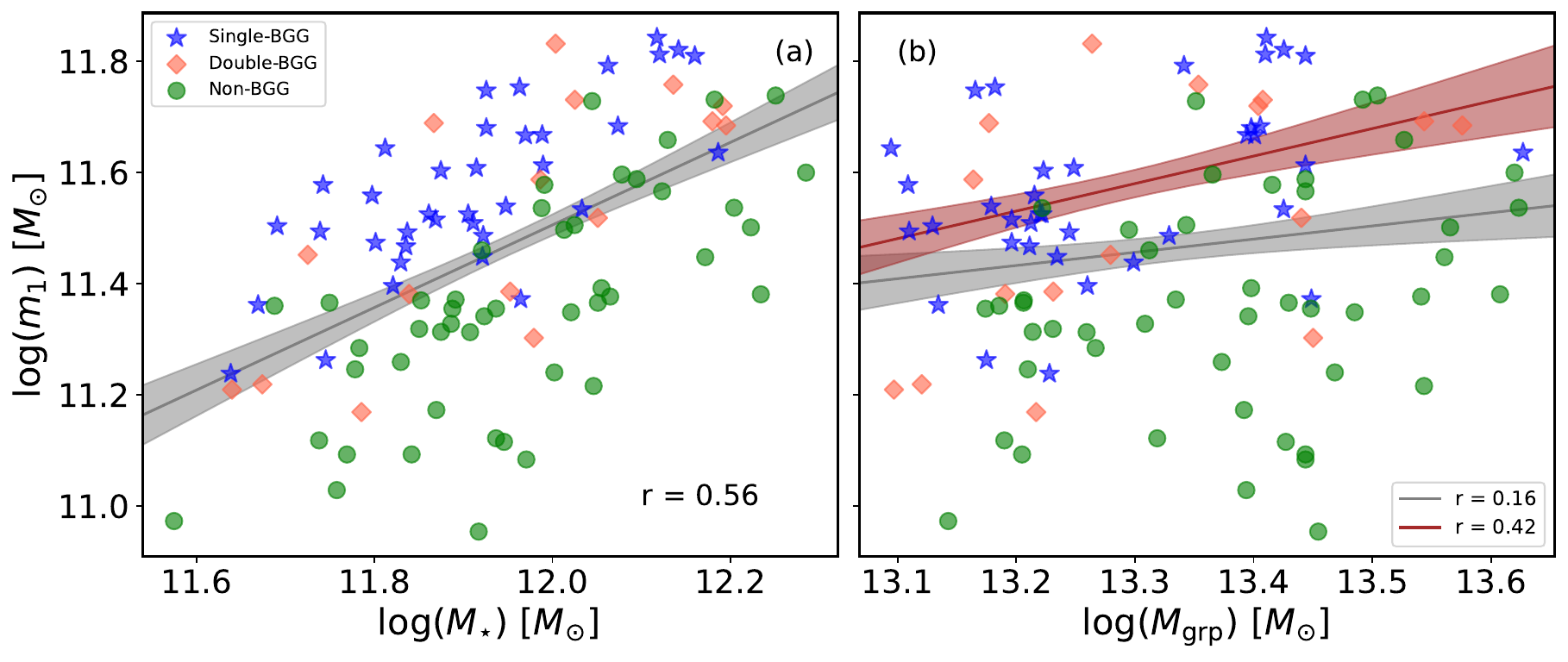}
    \caption{Same as Fig.~\ref{fig:Mvisib_mICL_bar_figure}, but for $m_{1}$ against (a) the total stellar mass ($M_\star$) and (b) the total system mass ($M_{\rm{grp}}$). The additional brown solid line and shaded band in panel (b) represent the linear fit and its associated uncertainty derived from the moderately strong positive linear correlation ($r = 0.42$) between $\log(m_{1})$ and $\log(M_{\rm{grp}})$, calculated after excluding non-BGG groups.}
    \label{fig:mBCG_mvis_linfit}
\end{figure*}

\subsubsection{Mass of first-ranked galaxies versus other group properties} \label{sec:m1vsprops}
Similar to the IGL mass, $m_{\rm{1}}$ exhibits a very high likelihood of linear correlation in logarithmic space ($p\geq 0.94$) with nearly all group parameters examined, except for the group scale radius, $\bar{R}_{\rm{gal}}$, where $p$ drops to $0.68$ and the Pearson's correlation coefficient $r$ approaches zero (see Table~\ref{tab:statistics}). The two strongest correlations of $m_{\rm{1}}$ are with the total mass-to-stellar mass ratio ($r=-0.70$) and with the total luminous mass ($r=0.56$; see also Fig.~\ref{fig:mBCG_mvis_linfit}a). Conversely, we find a practically insignificant positive correlation ($r=0.16$) between the logarithms of $m_{\rm{1}}$ and $M_{\rm{grp}}$, which appears to conflict with recent studies suggesting a stronger relationship \citep[e.g.][]{2020ApJ...891..129S}. This lack of correlation could be attributed either to the relatively narrow range of total group masses explored in our simulations or, more likely, to the highly unrelaxed state of the systems under investigation. In fact, when we exclude from these calculations the groups that apparently are less dynamically evolved, i.e.\ those lacking a clear BGG, the linear correlation between the logarithms of $m_{\rm{1}}$ and $M_{\rm{grp}}$ strengthens to a moderate level ($r=0.42$), and exhibits a slope $a=0.55$ indicative of a positive sublinear relationship between these two variables in linear space (see Fig.~\ref{fig:mBCG_mvis_linfit}b). The two panels of Fig.~\ref{fig:mBCG_mvis_linfit} also show that, for groups with the same total mass, whether stellar or global, first-ranked galaxies are systematically less massive in systems without a clear BGG.

\subsection{Quantifying the agreement between IGL and total mass density distributions} \label{sec:profiles}
Recent observational studies of diffuse light in clusters \citep{2010ApJ...717..420J, 2019MNRAS.482.2838M, 2020ApJ...901..128C, 2022ApJS..261...28Y} have gathered growing evidence that diffuse intergalactic light serves as a reliable tracer of the total mass distribution in galaxy systems, regardless of their dynamical state. To assess the robustness of this finding and identify potential limitations, we have generated and compared the radial distributions of the azimuthally averaged surface density profiles for both the IGL and the total mass in each one of the three orthogonal projections of our simulated groups at $z=0$. 
\begin{figure}
\centering
	\includegraphics[width=7cm, height=5cm]{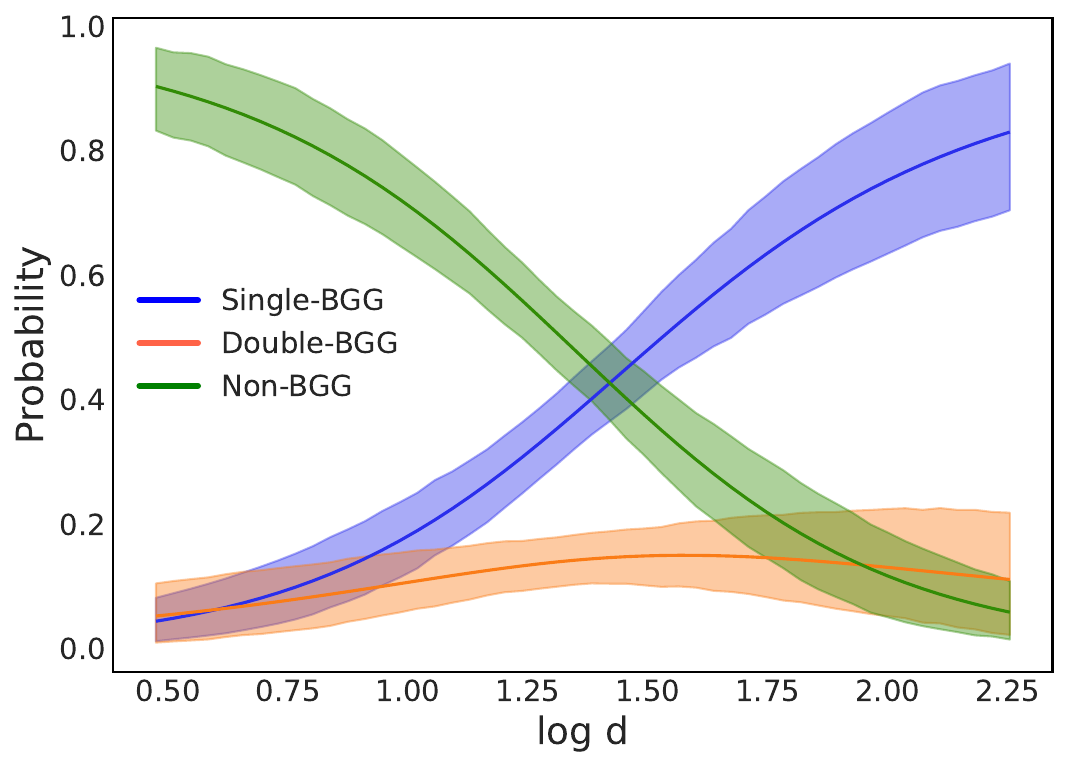}
    \caption{Probability that the metric $d$ measuring the fractional distance between the radial surface density profiles of total mass and IGL takes on a certain value by BGG class.}
    \label{fig:D_value}
\end{figure}
The radial surface density profiles were computed by dividing the particles into 25 circular rings, spanning from the groups' centre of mass to the maximum projected group-centric radius reached by the diffuse light. The bin boundaries were determined using a geometric sequence, ensuring a progressively decreasing number of particles per bin, with the final bin containing $10\%$ of the particles in the first bin. This approach yields relatively uniform bin spacing, as long as the densest central regions and most diluted outer regions of the groups are excluded. Specifically, we omit the $5\%$ of IGL particles nearest to and farthest from the centre, which defines the minimum ($R_{\rm min}$) and maximum ($R_{\rm max}$) radii of the density profiles for both the IGL and the total system mass. Therefore, we have:
\begin{equation}\label{r}
a_{n} = a_{1} \eta^{n-1} = a_{1} \gamma^\frac{n-1}{N_{\rm bin}-1}\;, \ \ \ \ \ \ \ \ \ \ \ \ \ \ \ \ \ \ \ \ \ \ \ \ n = 1,\ldots,N_{\rm bin}
\end{equation}
where $a_n$ is the number of particles of each component in the $n^{\rm th}$ bin of the sequence, $N_{\rm bin}$ is the total number of bins (25), and $\eta < 1$ is the common ratio defined from the adopted fraction of particles between the last bin and the first one, $\gamma=0.1$. The fact that the total number of particles between $R_{\rm min}$ and $R_{\rm max}$ must satisfy the formula for the sum of a finite geometric sequence:
\begin{equation}\label{npart}
S_{N_{\rm bin}} = a_1\left(\frac{1-\eta^{N_{\rm bin}}}{1-\eta}\right)\;, \ \ 
\end{equation}
is used to determine the number of particles $a_1$ in the first bin and to subsequently calculate the number of particles $a_n$ in the remaining circular bins from eq.~(\ref{r}). The surface density profiles are obtained by linearly interpolating the measured surface densities in each bin.    

The discrepancy between the surface density profiles of the total system mass and IGL mass for individual groups has been quantified through a metric inspired by those used in shape-matching 
\begin{equation}\label{d}
d(X,Y) = 100\times\mbox{med}\left(\left|\frac{\log Y^*(R_b) - \log X(R_b)}{\log X(R_b)}\right|\right) \;,   
\end{equation}
where the $X(R_b)$ and $Y(R_b)$ subsets represent, respectively, the binned radial profiles of the total and IGL mass, and $R_b$ is the projected group-centric bin radius. The asterisk on the tracer surface density profile indicates that this profile has to be renormalized so that its integral matches that of the total mass profile of the system within the region of the group where reliable measurements of both profiles can be made. An interesting property of this 'distance' is that it provides a direct quantification of the fractional difference between the shapes of the datasets that are being compared. Besides, the use of the median of the local distances makes the outcome fully robust to outliers. 

\begin{figure*}
\centering
	\includegraphics[width=17cm, height=8.5cm]{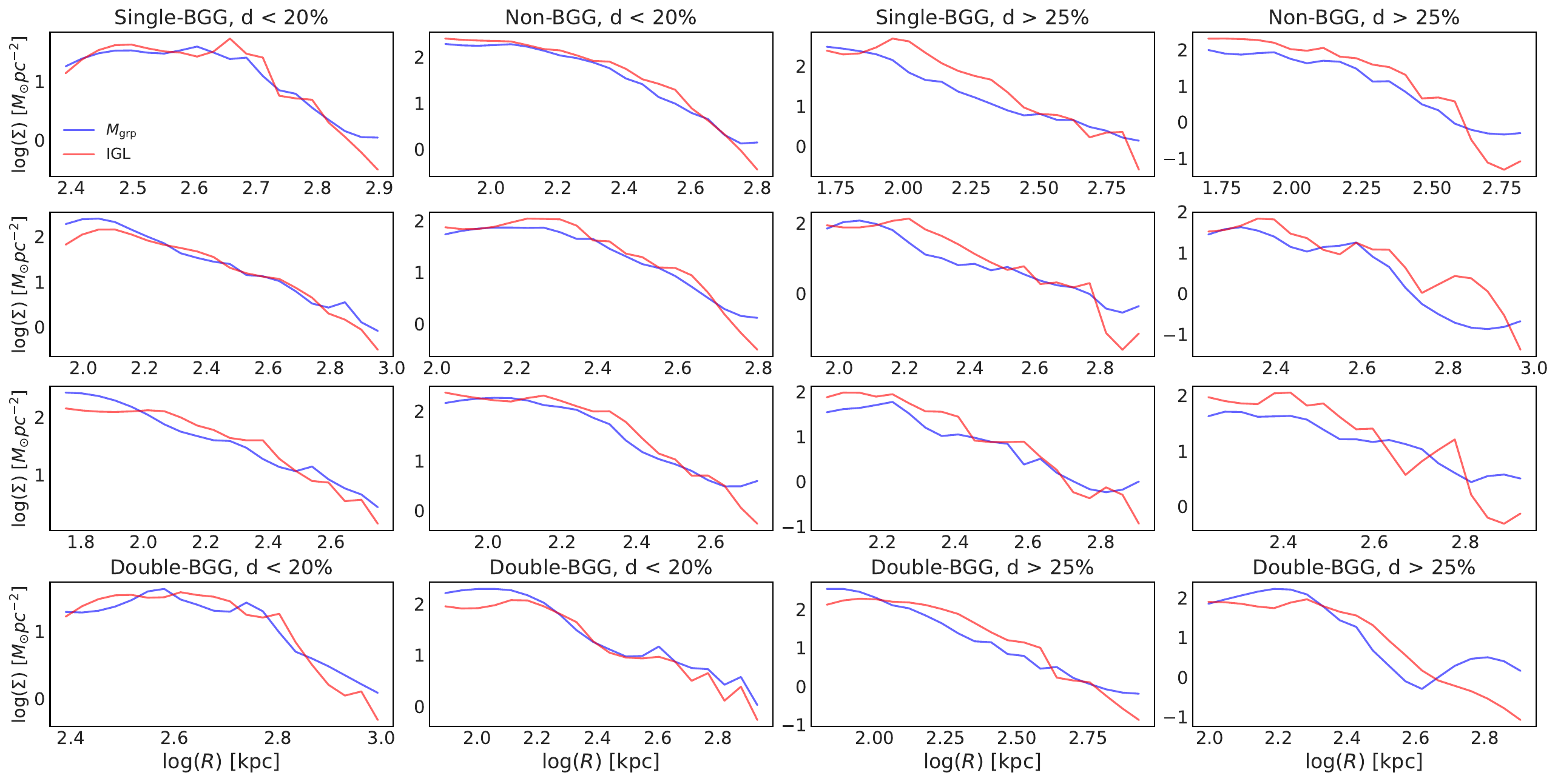}
    \caption{Radial surface density profiles for the IGL (red) and total system mass (blue) from one orthogonal projection of 16 individual groups, categorized as either single-BGG, double-BGG or non-BGG. The first two columns on the left display groups with $d < 20\%$, while the two columns on the right correspond to groups with $d > 25\%$, except for the last row, which presents double-BGG groups obeying these same constraints on $d$.}
    \label{fig:IGL_DM_profile}
\end{figure*}
\begin{figure*}
\centering
	\includegraphics[width=5cm, height=6cm]{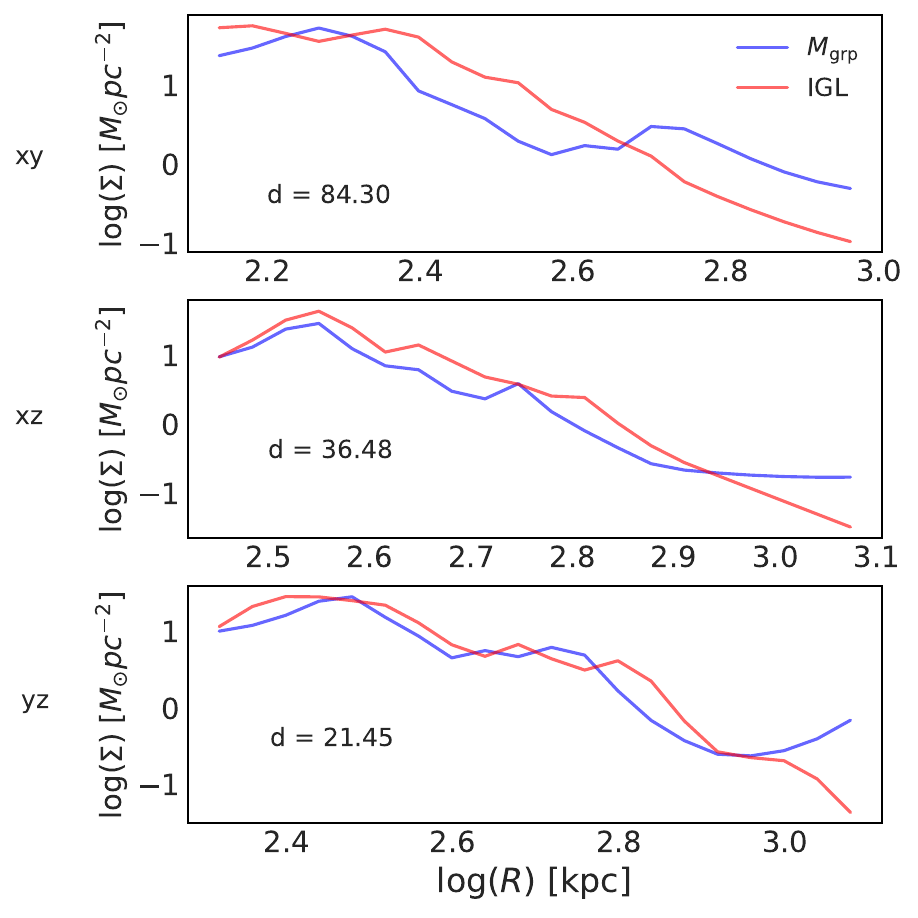}
    \includegraphics[width=4.3cm, height=6cm]{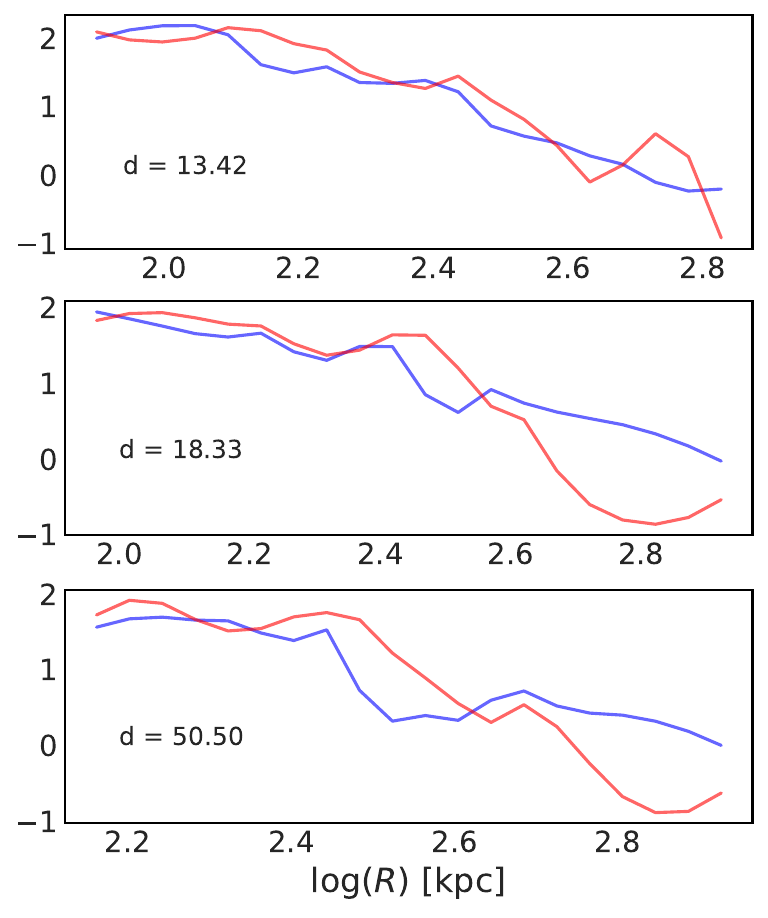}
    \includegraphics[width=4.3cm, height=6cm]{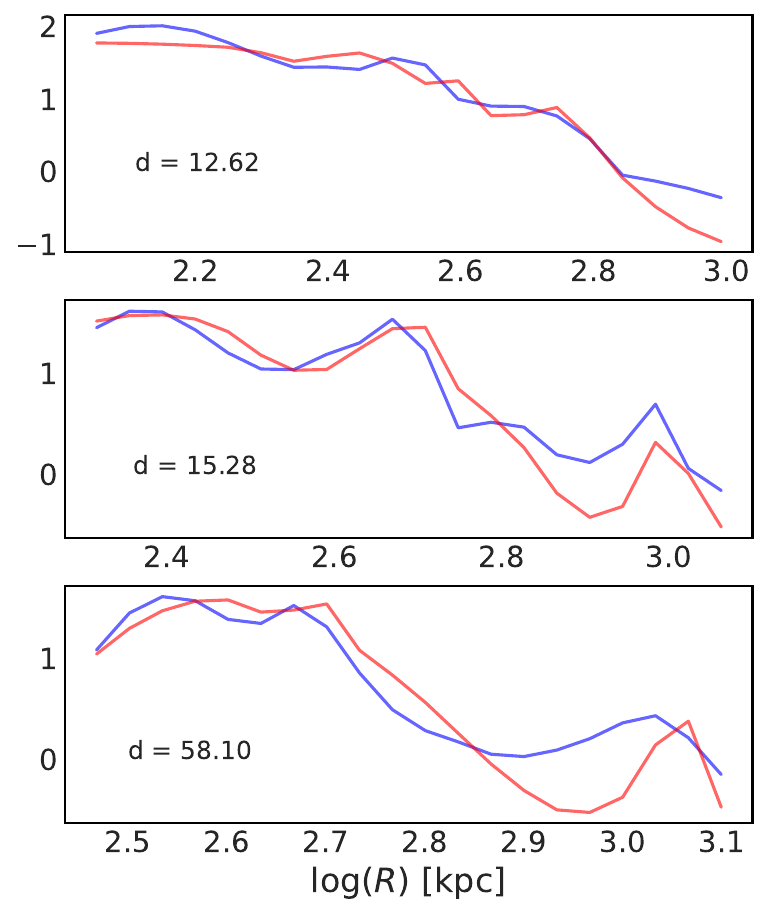}
    \includegraphics[width=4.3cm, height=6cm]{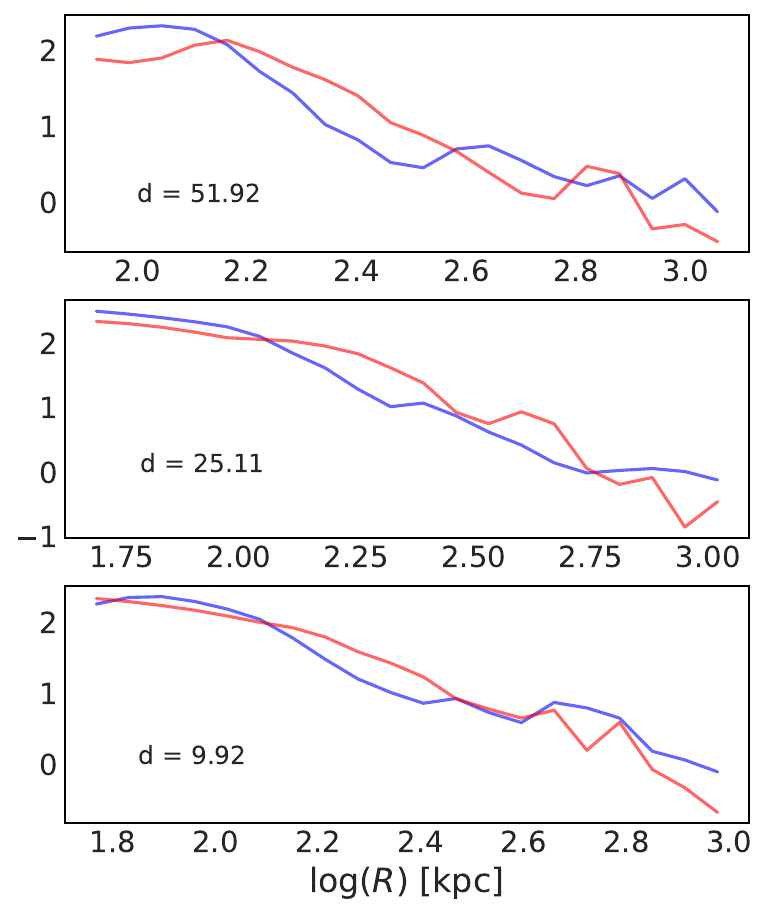}
    \caption{Total system mass and IGL surface density profiles of four different groups (from left to right) along the three orthogonal projections of their respective $z=0$ snapshots. The $d$-value measuring the 'percentage difference' between both profiles (see text) is included in each panel.}
    \label{fig:profile_projection}
\end{figure*}

The median (Q2) of the distribution of distances between surface density profiles arising from the three orthogonal projections of our simulated groups is $21\%$, with $Q1 = 12.5\%$ and $Q3 = 31.9\%$. Overall, two-thirds of the groups exhibit notable similarities between the radial surface density profiles of the IGL and the total system mass, with $d < 25\%$. Based on their BGG classification, non-BGG groups generally show smaller distances compared to single-BGG groups, whereas double-BGG systems, as observed in previously analysed parameters, display a relatively neutral trend, with only a slight tendency toward larger $d$ values (see Fig.~\ref{fig:D_value}). Examples of individual group profiles with $d$ values below $20\%$ and above $25\%$ are presented in Fig.~\ref{fig:IGL_DM_profile}. In Fig.~\ref{fig:profile_projection}, we compare the total system mass and IGL surface density profiles of four groups across the three orthogonal projections of their $z=0$ snapshots. This figure highlights that viewing direction affects not only the values of $d$ but also the shape of the profiles, reflecting the typically anisotropic distribution of matter in these systems.

\begin{figure*}
\centering
	\includegraphics[width=14cm, height=5.5cm]{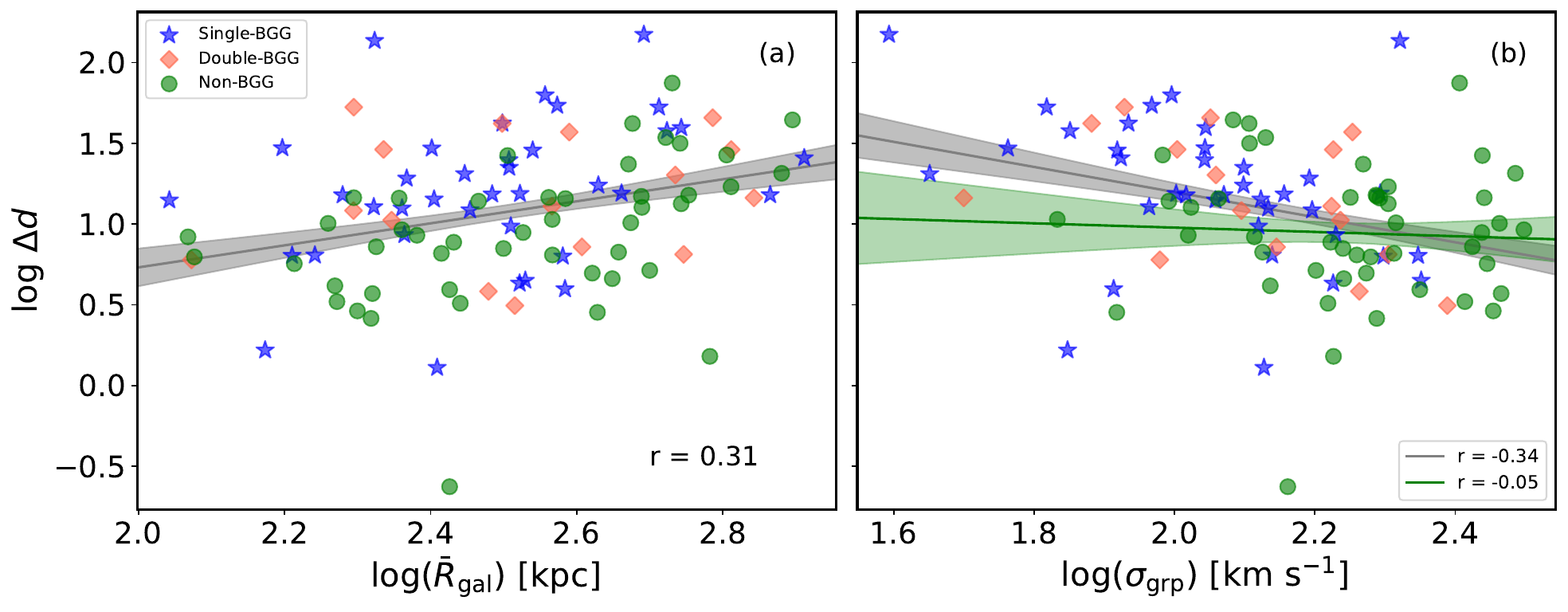}
    \caption{Same as Fig.~\ref{fig:Mvisib_mICL_bar_figure}, but for the correlation of $\Delta d$ with (a) the mean galaxy separation ($\bar R_{\mathrm{gal}}$) and (b) the group internal velocity dispersion ($\sigma_{\rm{grp}}$). The additional linear fit in panel (b) represented by the green solid line and the shaded band is drawn to illustrate the virtually null linear correlation ($r = -0.05$) that exists between $\log(\Delta d)$ and $\log(\sigma_{\rm{grp}})$ for non-BGG groups.} 
    \label{fig:deltaD}
\end{figure*}

We have also measured the difference between the maximum and minimum $d$-values, $\Delta d$, inferred from the three orthogonal projections of each group to explicitly account for the effects of projection on the fractional differences between the surface density profiles. As Fig.~\ref{fig:deltaD} shows, this parameter, which is positively correlated with $d$, has a weak positive correlation with the mean galaxy separation ($r = 0.31$) and a weak negative correlation with the group velocity dispersion ($r = -0.34$), and by extension, with the total system mass. In addition, when groups are categorized by their BGG class, we find that the latter correlation holds only for groups hosting a BGG, as for non-BGG groups the Pearson's correlation coefficient becomes basically null ($r = -0.05$); see panel b. This figure further illustrates that, while the similarity between the surface density profiles of the IGL and total mass depends on viewing direction, projection-related differences in $d$ remain relatively moderate, averaging between $\sim 5$–$25\%$ across the parameter range explored, with non-BGG groups consistently exhibiting lower $\Delta d$ values, which rarely exceed $25\%$, compared to their single-BGG counterparts.

\section{Summary and Conclusions} \label{sec:conclusions}
In this work, we conducted 100 cosmologically consistent, controlled numerical simulations of low-mass ($\sim\!1$--$5\,\times\,10^{13}\,M_\odot$) galaxy groups, tracing their early formation from $z=3$, when they behave as expanding overdensities, through the completion of their initial non-linear gravitational collapse at $z=0$. These simulations aim to investigate the production of ex-situ diffuse intergalactic light in these common, dense galaxy aggregates, generated by the strong gravitational interactions among member galaxies that take place during the early stages of group assembly. To distinguish galactic from intergalactic starlight, we applied the widely used technique of setting an upper surface brightness limit for the latter. In our simulations, this was achieved by imposing a cutoff in the projected stellar density corresponding to a surface brightness level of $26.5$ mag arcsec$^{-2}$ in the $V$ band, a threshold commonly adopted in the literature. After identifying the IGL, we examined the influence of various group parameters on its formation by looking for correlations between them and the fraction and total mass of this component, as well as the role played by the brightest group galaxies in this whole process. Notably, one of the key innovation of this study is the discovery that the trends in IGL properties are closely tied to the presence or absence of a dominant galaxy. By subdividing our group sample based on whether a BGG was present at the end of our simulations, we were able to uncover subtle nuances that would have otherwise been obscured when considering the group sample as a whole. Additionally, we explored the potential of using the IGL as a luminous tracer for the total matter distribution in galaxy aggregations, like the ones studied here, that are far from dynamical equilibrium. A summary of our main results follows:

\begin{itemize}
    \item The IGL begins to form in substantial amounts after the turnaround epoch, which in our simulations occurs at a median redshift of $\sim 0.85$, and keeps growing steadily with cosmic time until galaxy groups reach maximum compression around the end of the simulations at $z = 0$. The ex-situ IGL fractions produced by our simulated groups in this stage range from $\sim 6$ to $18\%$, with a median value of $11.3\%$.
    \item Examination of the fractional abundances across the different BGG classes of the groups, in relation to both various physical properties (total system mass, total stellar mass, velocity dispersion, mean projected intergalactic separation, and IGL fraction), and the fractional distance between the radial surface density profiles of total mass and IGL, reveals consistently opposing behaviours between single-BGG groups and non-BGG systems for all parameters analysed. Double-BGG systems, by contrast, exhibit a rather neutral behaviour, with minimal variations in their relative abundance within the ranges of parameter values covered by the simulations.
    \item The mass of IGL is strongly positively correlated with the total stellar mass of the group ($r = 0.87$), following an essentially linear relationship. However, the IGL fraction is completely independent of this property.
    \item While our findings suggest that the growth of the IGL parallels the development of the first-ranked group galaxy, its formation is not solely tied to the build-up of a dominant BGG, as significant amounts of diffuse intergalactic light are also observed in some non-BGG groups.
    \item The weak negative correlation between the IGL fraction and the internal velocity dispersion shown by our groups supports previous findings that low-velocity gravitational encounters between member galaxies enhance IGL formation by increasing the effectiveness of interactions.
    \item Our simulations confirm that the IGL serves as a reliable tracer of the total gravitational potential in galaxy groups, even when such systems are far from dynamical equilibrium. In two-thirds of our runs the radial surface density profiles of the IGL and total mass align closely, with fractional discrepancies below $25\%$. This agreement persists across groups of all types, but is most noticeable in those lacking a BGG, as well as in more compact, massive, and high-velocity dispersion systems. However, our experiments also reveal that the degree of alignment is affected by projection effects and the spatial anisotropy inherent to galaxy aggregations, particularly in dynamically unrelaxed systems, therefore underscoring its dependence on the viewing direction.
\end{itemize}

\begin{acknowledgements}

The authors express their gratitude to the anonymous referee for insightful comments and constructive feedback, which have greatly contributed to improving the quality and clarity of this work. We also extend our thanks to Yolanda Jim\'enez-Teja for engaging discussions on some aspects of this study. Financial support from the Spanish state agency MCIN/AEI/10.13039/501100011033 and by 'ERDF A way of making Europe' funds through research grants PID2022-140871NB-C21 and PID2022-140871NB-C22 is also acknowledged. This work has been also partially supported by MCIN/AEI/10.13039/501100011033 through the Centre of Excellence Severo Ochoa's award for the Instituto de Astrof\'\i sica de Andalucía under contract CEX2021-001131-S and the Centre of Excellence Mar\'\i a de Maeztu's award for the Institut de Ci\`encies del Cosmos at the Universitat de Barcelona under contract CEX2019–000918–M. BBW acknowledges funding from the pre-doctoral fellowship PRE2020-093715 associated with the SEV–2017–0709 project. 
\end{acknowledgements}

\bibliographystyle{aa} 
\bibliography{biblio}


%
%

\end{document}